\documentclass[pra,aps,footinbib,twocolumn,superscriptaddress,citeautoscript,longbibliography]{revtex4-1}

\usepackage[utf8x]{inputenc}
\usepackage{array}
\usepackage{amssymb}
\usepackage{amsmath}
\usepackage{amsfonts}
\usepackage{color}
\usepackage{graphicx}
\usepackage{bm}
\usepackage{dsfont}
\usepackage{url}
\usepackage{braket}
\usepackage{empheq}
\usepackage{tikz}
\usepackage[unicode]{hyperref}
\usepackage{color}
\usepackage{soul}

\hypersetup{
   unicode=true,          
   plainpages=false,
   colorlinks=true,       
   citecolor=blue,        
}
\allowdisplaybreaks
\graphicspath{{.}{./fig/}}

\newcommand{\field}[1]{\mathbb{#1}} 

\newcommand{\rme}{\mathrm{e}}
\newcommand{\rmi}{\mathrm{i}}
\newcommand{\x}{\mathrm{x}}
\newcommand{\y}{\mathrm{y}}

\renewcommand{\d}{\ensuremath{\mathrm{d}}}
\renewcommand{\u}{\mathbf{u}}

\newcommand{\sect}[1]{\section{#1} }
\newcommand{\subsect}[1]{\subsection{#1} }

\begin{document}

\title{Disorder-induced two-body localised state in interacting quantum walks}
\author{L. A. Toikka}
\email{lauri.toikka@gmail.com}
\affiliation{Institute for Theoretical Physics, University of Innsbruck, A-6020 Innsbruck, Austria}
\date{\today}

\begin{abstract}
We observe the onset of non-ergodicity from ballistic propagation of a two-body bound state in an interacting discrete-time quantum walk (DTQW) due to time-dependent disorder in the interaction. The effect of the disorder on the two-body state can be interpreted as Anderson-type localisation in a projected DTQW, but without saturation of entanglement entropy. We characterise the two-body localisation in terms of the rate of growth of entanglement entropy and compute the localisation length. We find indications for two distinct growth laws for the entanglement: a logarithmic law in the delocalised phase and a double logarithmic law in the localised phase. We discuss similarities with many-body localisation (MBL).
\end{abstract}

\maketitle

\sect{Introduction}
Classical random walks have been succesful at modelling a wide range of phenomena across the natural sciences, finance~\cite{doi:10.1086/260062}, and computer science~\cite{norris1998markov}. Originally proposed by Feynman~\cite{FeynmanHibbs1965} and Aharonov \textit{et al}.~\cite{PhysRevA.48.1687}, quantisation of a discrete random walk results in coherent motion of a quantum particle on a graph~\cite{Venegas-Andraca2012}. The coin that decides the direction of a step becomes a qubit such that after the first step the particle is in a superposition determined by the quantum state of the coin. If the quantum particle is measured at each step or generally if there is decoherence, the probability distribution reduces to a classical random walk; otherwise, the particle can interfere with itself in a coherent way and quantum correlations develop in the system. Quantum parallellism alone is not sufficient for useful quantum algorithms, but combined with the interference enable quantum walks, both continuous and discrete-time, to form a promising primitive for quantum algorithms and quantum simulation~\cite{PhysRevLett.102.180501,PhysRevA.81.042330}, to be run naturally and efficiently on quantum computers. 

Statistical algorithms based on random walks often represent some of the simplest and fastest known ways to solve hard problems, for example, underpinning the fastest known tests for primality~\cite{RABIN1980128} and Monte Carlo methods~\cite{newman1999monte}. Discrete-time quantum walks (DTQWs) find many applications as simulators of condensed matter physics including disorder and single-particle Anderson localisation~\cite{PhysRevA.94.023601,PhysRevE.82.031122,PhysRevA.92.052311,PhysRevB.96.144204}, and are able to represent all the known classes of topological phases in one and two dimensions~\cite{PhysRevA.82.033429}. Shenvi \textit{et al.}~\cite{PhysRevA.67.052307} showed that a DTQW can reproduce the optimal $\sqrt{N}$ quantum speedup in search times found earlier with Grover’s algorithm~\cite{PhysRevLett.79.325} for finding a flagged entry in an unsorted database with $N$ entries. Generally, the unitary walk operator $S$ can be thought of as a Floquet system, that is, as a stroboscopic simulator $\rme^{-\rmi H}$ for some effective Hamiltonian $H$~\cite{PhysRevB.88.121406}. In this sense, a DTQW is a versatile platform for simulating Hamiltonian dynamics in wide-ranging lattice systems.

DTQWs have been experimentally realised on platforms ranging from trapped ions~\cite{PhysRevLett.103.090504}, trapped atoms~\cite{Karski174}, and correlated photons~\cite{Peruzzo1500} to nuclear magnetic resonance~\cite{PhysRevA.67.042316}. The experimental implementation of a DTQW with trapped atoms in an optical lattice~\cite{Karski174} involves quite unavoidably interactions when two atoms occupy the same site in the lattice. Depending on the duration the atoms are in contact these interactions take the form of a multiplicative phase factor, which is not dissimilar from the interchange of two Abelian anyons that can also be thought of as a collisional phase interaction~\cite{KITAEV20062}. Anyon fusion then resembles the formation of a molecular dimer, decaying exponentially in the relative position of the two atoms, which has been theoretically predicted and shown to undergo a slower DTQW in its own right~\cite{Ahlbrecht_2012}. In addition to the entanglement between the quantum state (e.g. spin) and position of the particles generated by the quantum coin and subsequent shift operation, constituting a step in the single-particle DTQW, the collisional phase interaction also entangles positions of the two particles. Therefore, the interaction amounts to an entangling quantum gate such that the interacting DTQW develops all-to-all entanglement between position and spin degrees of freedom of all the particles. An entangling gate is an essential resource for universal gate-based quantum applications since a controlled NOT gate, responsible for creating entanglement, together with single-qubit rotations forms a complete set~\cite{nielsen2010quantum}.

Here, we study the effect of temporal disorder in the interaction on the DTQW dynamics. This type of disorder can be expected to arise from fluctuations in the experimental control of the DTQW sequence as the phase is strongly sensitive to the duration of contact. While it seems reasonable to intuitively expect that disorder in the interaction would generally result in decoherence, moving and obfuscating information about the initial state making it later accessible to only highly non-local observables which would collapse the quantum coherence properties of the walk, we find that the two-walker DTQW instead becomes exponentially localised. Instead, the disordered dynamics preserves information and direct memory about the two-body initial state. We show that the localisation can be understood in terms of Anderson-type localisation by using the transfer matrix approach, but with different entanglement properties. We compute the localisation length, and find two distinct growth laws for entanglement: a logarithmic law in the delocalised phase and a double logarithmic law in the localised phase. Finally, we discuss similarities with many-body localisation (MBL)~\cite{BASKO20061126,doi:10.1146/annurev-conmatphys-031214-014726}.

The paper is organised as follows. We present the physical system, compute the transfer matrix and band structure of the ordered limit in Sec.~\ref{sec:2}. Numerical results for the growth of entanglement and localisation length are shown in Secs.~\ref{sec:3a} and~\ref{sec:3b} respectively. We conclude and discuss the outlook in Sec.~\ref{sec:dc}.

\sect{\label{sec:2}Physical system}
In a classical discrete random walk on a line, at each step of the walk a particle moves left or right according to the result of tossing a coin. In the quantum case we consider here, the single-particle degrees of freedom are the position of the particle in a one-dimensional chain with site $i$ at $x_i$, and the two-level internal state of the particle, $\zeta = \pm 1$. We denote the single-particle basis states as $\ket{X} \otimes \ket{\zeta} \equiv \ket{X,\zeta}$, where $\ket{\zeta} = c_+ \ket{+}+ c_- \ket{-}$ with $|c_+|^2 + |c_-|^2 = 1$, and $\ket{X} = \sum_i d_i \ket{x_i} $ with $\sum_i |d_i|^2 = 1$ and $x_i$ enumerating the positions in the lattice. With these definitions, the state-dependent shift is given by the operator
\begin{equation}
\label{eqn:G}
G = \sum_{i \in \field{Z}} {\left(\ket{x_{i + 1}}\bra{x_i} \otimes \ket{+}\bra{+} + \ket{x_{i - 1}}\bra{x_i} \otimes \ket{-}\bra{-} \right) },
\end{equation}
which moves the particle to the right (left) if the internal state has support in $\ket{+}$ ($\ket{-}$). Quite generally, the spin state of the particle can be manipulated on the Bloch sphere. With a walk on a line, in the basis $\ket{-} = (1,0)^\mathrm{T}$, $\ket{+} = (0,1)^\mathrm{T}$, the most general coin toss transforming the internal state is given by~\cite{DBLP:journals/jcss/BachCGJW04}
\begin{equation}
\label{eqn:coin}
C = \begin{pmatrix}
\sqrt{\rho} & \sqrt{1-\rho} \\
\sqrt{1-\rho} & -\sqrt{\rho} 
\end{pmatrix} ,
\end{equation}
with the position on the Bloch sphere of the initial qubit,
\begin{equation}
\ket{\zeta(0)} = \eta \ket{+}+ \sqrt{\eta - 1} \rme^{\rmi \beta} \ket{-},
\end{equation}
being fully characterised by the two real parameters $\eta$ and $\beta$. We focus on the Hadamard coin meaning we take $\rho = 1/2$ (an unbiased coin). The total unitary operator progressing the single-particle walk by one step, $S$, is then given by
\begin{equation}
S = G\left(\mathrm{Id}_\mathbb{Z} \otimes C \right),
\end{equation}
where the identity on the position space reflects the fact that the position of the particle is not modified during the coin toss itself. The state after $t$ steps is given by
\begin{equation}
\label{eqn:St}
\ket{\Psi(t)} = S^t \ket{\Psi(0)},
\end{equation}
where $\ket{\Psi(0)} = \ket{X(0)}\otimes\ket{\zeta(0)}$ is the initial state.

The Hilbert space of the single-particle walk is given by $\mathcal{H}^{(1)} = L^2(\field{Z}) \otimes \mathbb{C}^2$, and the two-particle Hilbert space follows as a tensor product,
\begin{equation}
\mathcal{H}^{(2)} = \mathcal{H}_1^{(1)} \otimes \mathcal{H}_2^{(1)},
\end{equation}
where we label the two particles by $1$ and $2$. The total unitary operator progressing the two-particle walk by one step, $S^{(2)}$, is given by
\begin{equation}
\label{eqn:S2def}
S^{(2)} = V^{(2)}\left(G_2^{(1)} C_2^{(1)} \otimes G_1^{(1)} C_1^{(1)} \right),
\end{equation}
where the operator $V^{(2)}$ represents the collisional phase interaction that adds a phase $\mu$ when the two particles exist at the same site. We have
\begin{equation}
\label{eqn:V2def}
\begin{split} 
V^{(2)} &= \sum_{\zeta_{i,j} \in \{\pm \}} \left\lbrace \sum_{i\in \mathbb{Z}}   \mathrm{e}^{\mathrm{i} \mu} \ket{ x_i \zeta_i x_i \zeta_j} \bra{x_i \zeta_i x_i \zeta_j} \right. \\
&\left.  \qquad + \sum_{(i, j) \in \mathbb{Z}^2; i \neq j } \ket{x_i \zeta_i x_j \zeta_j}\bra{x_i \zeta_i x_j \zeta_j}  \right\rbrace \\
&=\text{Id}_{\mathcal{H}^{(2)}} + \left(\mathrm{e}^{\mathrm{i} \mu} - 1\right)\sum_{i\in \mathbb{Z}}   \ket{x_i  x_i}\bra{x_i  x_i } \otimes \text{Id}_4,
\end{split}
\end{equation}
where $ \sum_{i\in \mathbb{Z}}   \ket{x_i  x_i}\bra{x_i  x_i } \otimes \text{Id}_4 \equiv \mathcal{P}$ acts as a projector onto the $X_1 - X_2 = 0$ manifold where the interaction takes place. The disorder appears when $\mu$ fluctuates between subsequent steps. We take $\mu$ to be uniformly distributed with a given variance around the mean value $\mu_0$.

The dynamics of the two-particle state $\ket{\Psi^{(2)}(t)}$ can be represented by the lattice-resolved four-component wavefunction
\begin{equation}
\Psi^{(2)}_{m n}(t) = \begin{pmatrix}
\Psi_{m n}^{+ +}(t) \\ \Psi_{m n}^{ + -}(t) \\ \Psi_{m n}^{ - +}(t) \\ \Psi_{m n}^{ - -}(t)
\end{pmatrix},
\end{equation}
defined at time $t$ on the two-dimensional lattice site with coordinates $m$ and $n$ for particles $1$ and $2$ respectively. Analogously with Eq.~\eqref{eqn:St}, the state after $t$ steps is given by
\begin{equation}
\label{eqn:St2}
\ket{\Psi^{(2)}(t)} = \left( S^{(2)}\right)^t \ket{\Psi^{(2)}(0)}.
\end{equation}

\subsect{Transfer matrix for disordered case}

To facilitate the study of two-body localisation in \textit{position space}, which here results from disorder in the \textit{time-dependent} collisional phase factor $\mu(t)$, we express Eq.~\eqref{eqn:St2} in terms of a `transfer matrix' -type formalism~\cite{crisanti2012products} that propagates the wavefunction in position space. The transfer matrix is useful because during one step of the DTQW the particles move only by one lattice site.

First, we note that the DTQW can be formally related to a Floquet eigenvalue equation
\begin{equation}
\label{eqn:Floquet}
S^{(2)}\ket{\phi} = \rme^{-\rmi \phi} \ket{\phi}
\end{equation}
because the spectrum of a unitary operator must lie on the unit circle. In analogy with standard Floquet theory~\cite{stockmann2006quantum} the energy can only be defined modulo $2\pi \hbar /\tau$; we scale $\hbar = \tau = 1$. For a localised mode we focus on eigenstates of $S^{(2)}$ meaning that from Eqs.~\eqref{eqn:St2} and~\eqref{eqn:Floquet} the unitary eigenvalue problem becomes
\begin{equation}
\label{eqn:Floquet1}
\ket{\Psi^{(2)}(t+1)}= \rme^{-\rmi \phi} \ket{\Psi^{(2)}(t)}.
\end{equation}
Without disorder, since the interaction depends only on the relative position of the particles, the total quasi-momentum $P$ is conserved. Here $P$ is an eigenvalue of the total shift operator $G_1^{(1)} \otimes G_2^{(1)}$ for the two particles. With disorder in $\mu(t)$, we can expect the two-body localised state wavefunction to decay exponentially inside the subspace $X_1 - X_2 = 0$. To simplify the treatment and to capture the essential features, we project the propagator $S^{(2)}$ onto the subspace $X_1 - X_2 = 0$. In other words, we can define an effective propagator
\begin{equation}
S_\mathcal{P}^{(2)} =\rme^{\rmi \mu} G \left(\mathrm{Id}_\mathbb{Z} \otimes C \right),
\end{equation}
which describes a single-particle DTQW in subspace $\mathcal{P}$ that comes with a phase factor $\mu$ at each step. In conjunction, we define the effective single-particle wavefunction: 
\begin{equation}
\braket{n|\phi(t)} \equiv \phi_{n}(t) = \begin{pmatrix}
\phi_{n}^{+}(t) \\ \phi_{n}^{-}(t)
\end{pmatrix},
\end{equation}
defined at time $t$ on a one-dimensional lattice site with coordinate $n$. A step in the effective single-particle DTQW is then defined by the equation 
\begin{equation}
\label{eqn:effDTQWTrEq}
\begin{split}
\rme^{-\rmi \mu_{n}} \ket{\phi_n(t+1)} &= \begin{pmatrix}
C_{11} & C_{12}\\
0 & 0
\end{pmatrix} \ket{\phi_{n-1}(t)} \\
&\qquad + \begin{pmatrix}
0 & 0\\
C_{21} & C_{22}
\end{pmatrix}  \ket{\phi_{n+1}(t)},
\end{split}
\end{equation}
where $C_{ij}$ are the matrix elements of the coin~\eqref{eqn:coin}.

Projecting Eq.~\eqref{eqn:Floquet1} with $\mathcal{P}$ to read $\ket{\phi(t+1)}= \rme^{-\rmi \phi} \ket{\phi(t)}$, we may now express the time-propagator $S_\mathcal{P}^{(2)}$ in terms of a transfer matrix $T_n$ propagating in position space as follows. For one step of the walk, from Eq.~\eqref{eqn:effDTQWTrEq} we thus have
\begin{subequations}
\label{eqn:treqsaux}
\begin{align} 
 \rme^{-\rmi \left(\mu_{n} + \phi\right)} \phi_{n}^{+}(t) &=C_{11} \phi_{n-1}^{+}(t) + C_{12} \phi_{n-1}^{-}(t) ,\\
 \rme^{-\rmi \left(\mu_{n} + \phi\right)}  \phi_{n}^{-}(t) &=C_{21} \phi_{n+1}^{+}(t) + C_{22} \phi_{n+1}^{-}(t).
\end{align}
\end{subequations}
Defining $\Phi_{n}(t) = \begin{pmatrix}
\phi_{n}^{+}(t) & \phi_{n-1}^{-}(t)
\end{pmatrix}^\mathrm{T}$, system~\eqref{eqn:treqsaux} is equivalent to a $2\times 2$ transfer matrix equation
\begin{equation}
\label{eqn:dMap}
\ket{\Phi_n(t)} = T_{n-1} \ket{\Phi_{n-1}(t)},
\end{equation}
where the transfer matrix is given by
\begin{equation}
\label{eqn:matTransfer}
T_{n} = \begin{pmatrix}
 \rme^{\rmi \left(\alpha_n + \phi\right)}\left(C_{11}  -\frac{C_{12} C_{21}}{C_{22} } \right)  &  \frac{ \rme^{\rmi \left(\alpha_n - \beta_n\right) }  C_{12}  }{C_{22} } \\
-\frac{ C_{21}}{C_{22} }  & \frac{ \rme^{-\rmi \left(\beta_n + \phi\right)}   }{C_{22} }
\end{pmatrix},
\end{equation}
where $\alpha_n = \mu_n$ and $\beta_n = \mu_{n-2}$ are random variables following some distribution. We take them uniformly distributed around $\mu_0$ with a given variance. If we restrict $\alpha_n = \beta_n \equiv \varepsilon_n$ and take $C_{12} \to \rme^{\rmi \gamma_n} C_{12}$, $C_{21} \to \rme^{-\rmi \gamma_n} C_{21}$, then in a conventional tight-binding band theory picture it is possible to interpret the diagonal disorder $\varepsilon_n$ as a potential energy, and the off-diagonal disorder $\gamma_n$ as a flux. Given the above physical interpretation, uncorrelated disorder in either can then be expected to result in Anderson localisation for the effective single-particle DTQW projected from the full two-body problem. Disorder in the coin itself can thus also be expected to result in Anderson localisation, but our focus here is on the interaction.

\begin{figure}[t]
  \centering
        \begin{tikzpicture}
        \def\x{4.5};        \def\y{3.0};
        \def\v{-1.1};   \def\vv{-1.3};;    \def\w{10.8};	     \def\u{10.8};
			 \node at (0.0*\x,0*\y) {    \includegraphics[width=0.45\textwidth,angle=0]{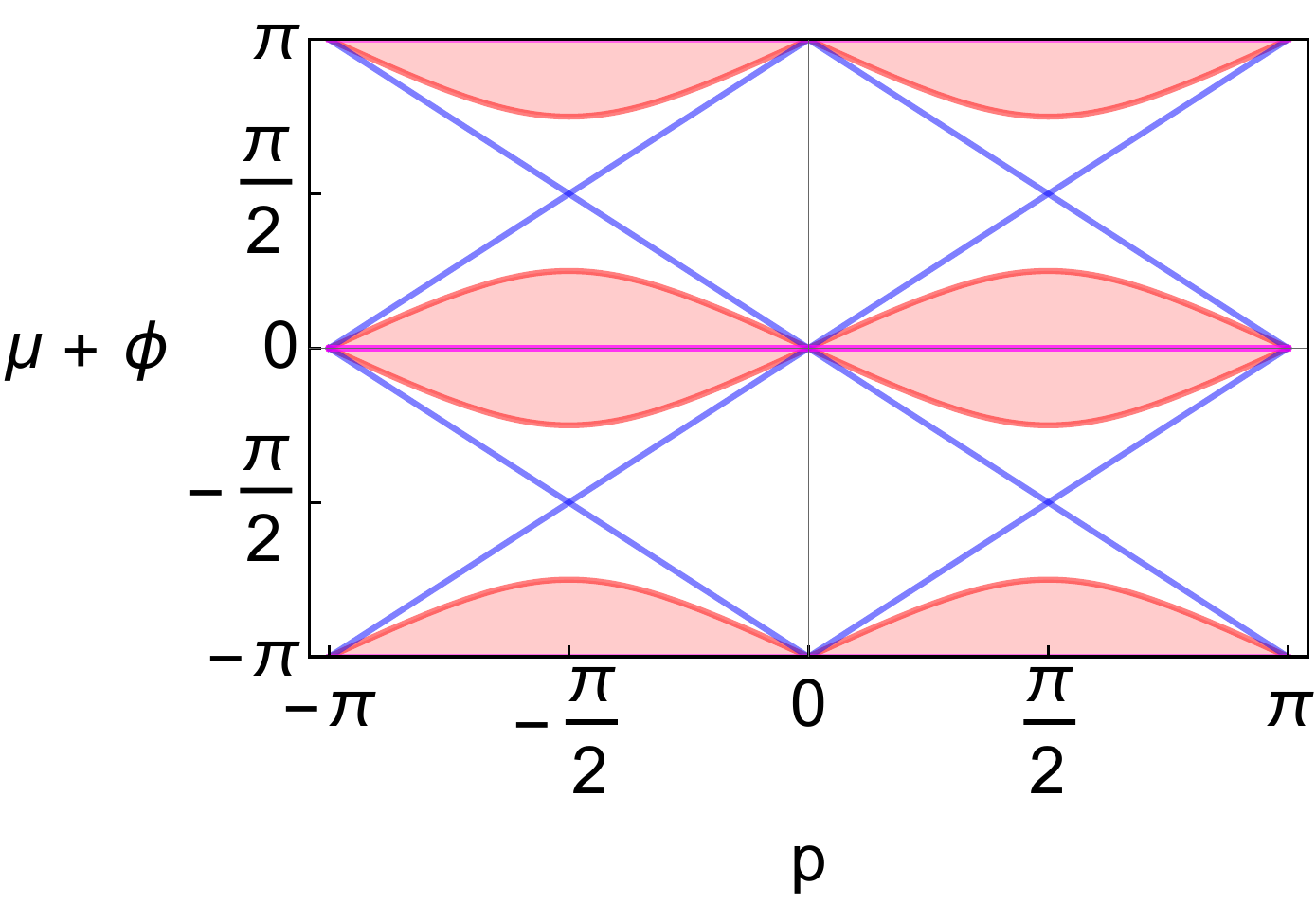}};		 
      \end{tikzpicture}
      
      \caption{ \label{fig:0} 
      \textit{Dispersion in the subspace $X_1 - X_2 = 0$ vs. quasi-momentum $p$.} The parameters correspond to the unbiased coin $\rho = 1/2$ (red), and fully biased coins $\rho = 1$ (blue, Dirac cones) and $\rho = 0$ (magenta, flat bands at $\mu + \phi = 0,\pm \pi$). 
      }   
\end{figure}

If the state $\Psi^{(2)} = \{\cdots, \Psi_{-n}^{(2)}, \cdots, \Psi_{0}^{(2)}, \cdots \Psi_{n}^{(2)}, \cdots \} $ is exponentially localised along the anti-diagonal (we denote $\Psi_{n}^{(2)} \equiv \Psi_{nn}^{(2)}$), then the wavefunction envelope decreases exponentially far from the maximum e.g. at $\Psi_{0}^{(2)}$,
\begin{equation}
\label{eqn:AndLoc}
\left| \Psi_{n}^{(2)} \right| \sim \left| \Psi_{0}^{(2)} \right| \rme^{-|n|/\xi}
\end{equation}
with the localisation length $\xi$. We compute the two-body localisation length $\xi$ by
\begin{equation}
\xi = \frac{1}{\Lambda},
\end{equation}
where $\Lambda$ is the Lyapunov exponent associated with the discrete map~\eqref{eqn:dMap}~\cite{crisanti2012products},
\begin{equation}
\label{eqn:HJT}
\Lambda =\lim_{x\to \infty} \frac{1}{x}  \ln{\left(\left|  \mathrm{Tr} \prod_{n=1}^x T_n \right|\right)} .
\end{equation}
The projected transfer matrix approach is a direct and fast method for computing the two-particle localisation length $\xi$. It is especially suitable for ensuring convergence by allowing to take $x \gtrsim 10^6$, whereas exact diagonalisation (or direct propagation) of the full two-body problem is not as feasible.

\subsect{Band structure in ordered case}

\begin{figure}[t]
  \centering
        \begin{tikzpicture}
        \def\x{4.5};        \def\y{3.0};
        \def\v{-1.1};   \def\vv{-1.3};;    \def\w{10.8};	     \def\u{10.8};
			 \node at (0.0*\x,0*\y) {    \includegraphics[width=0.45\textwidth,angle=0]{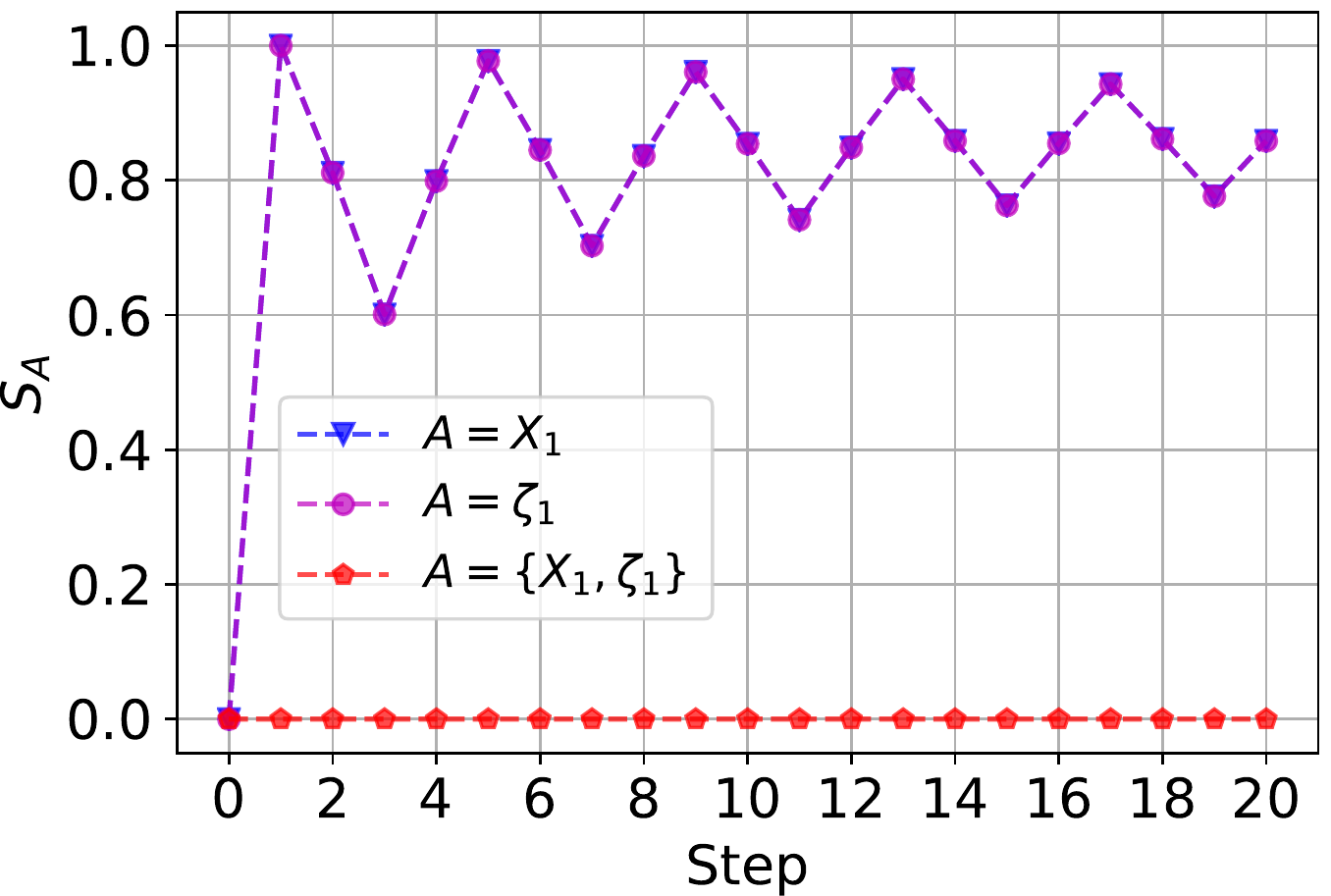}};
			 \node at (2.1,-0.2) {\includegraphics[width=0.19\textwidth,angle=0]{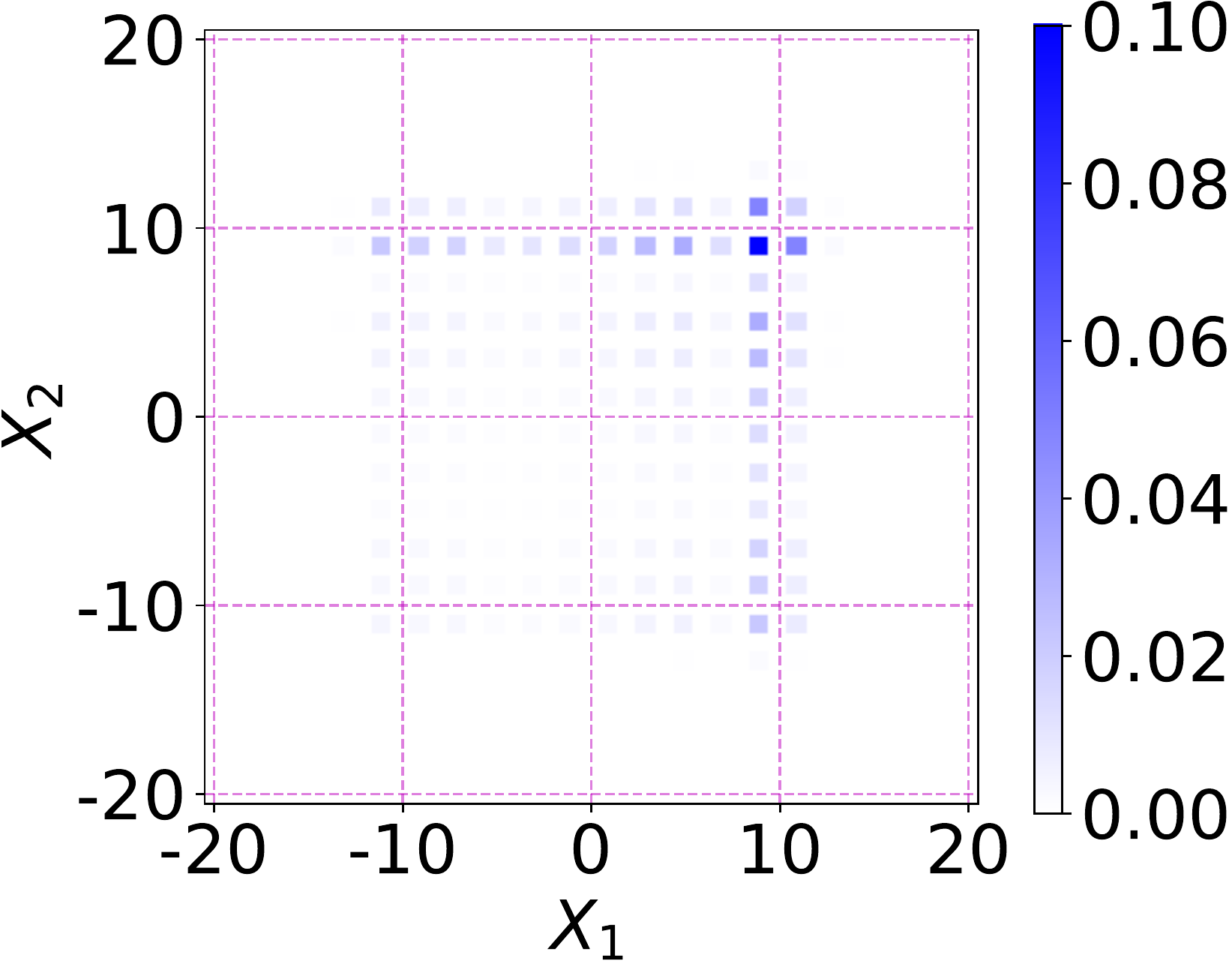}};

      \end{tikzpicture}
      
      \caption{ \label{fig:1} 
      \textit{Two non-interacting unbiased Hadamard walks ($\mu = 0$).} The entanglement entropy $S_A$ for various partitionings $A$, and probability distribution of the quantum walk at step 15 (inset).  Particles 1 and 2 remain unentangled (red line). We denote the position and spin of particle $i$ by $X_i$ and $\zeta_i$ respectively, and the total set of degrees of freedom is $X_1, \zeta_1, X_2, \zeta_2$. The initial condition, $\ket{\Psi(0)} = \ket{x_0} \otimes \ket{+}$, is the same for both particles. At odd (even) time steps the even (odd) sites are unoccupied, and the joint probability distribution is symmetric about the anti-diagonal.
      }   
\end{figure}

\begin{figure}[t]
  \centering
        \begin{tikzpicture}
        \def\x{4.5};        \def\y{3.0};
        \def\v{-1.1};   \def\vv{-1.3};;    \def\w{10.8};	     \def\u{10.8};
			 \node at (0.0*\x,0*\y) {    \includegraphics[width=0.45\textwidth,angle=0]{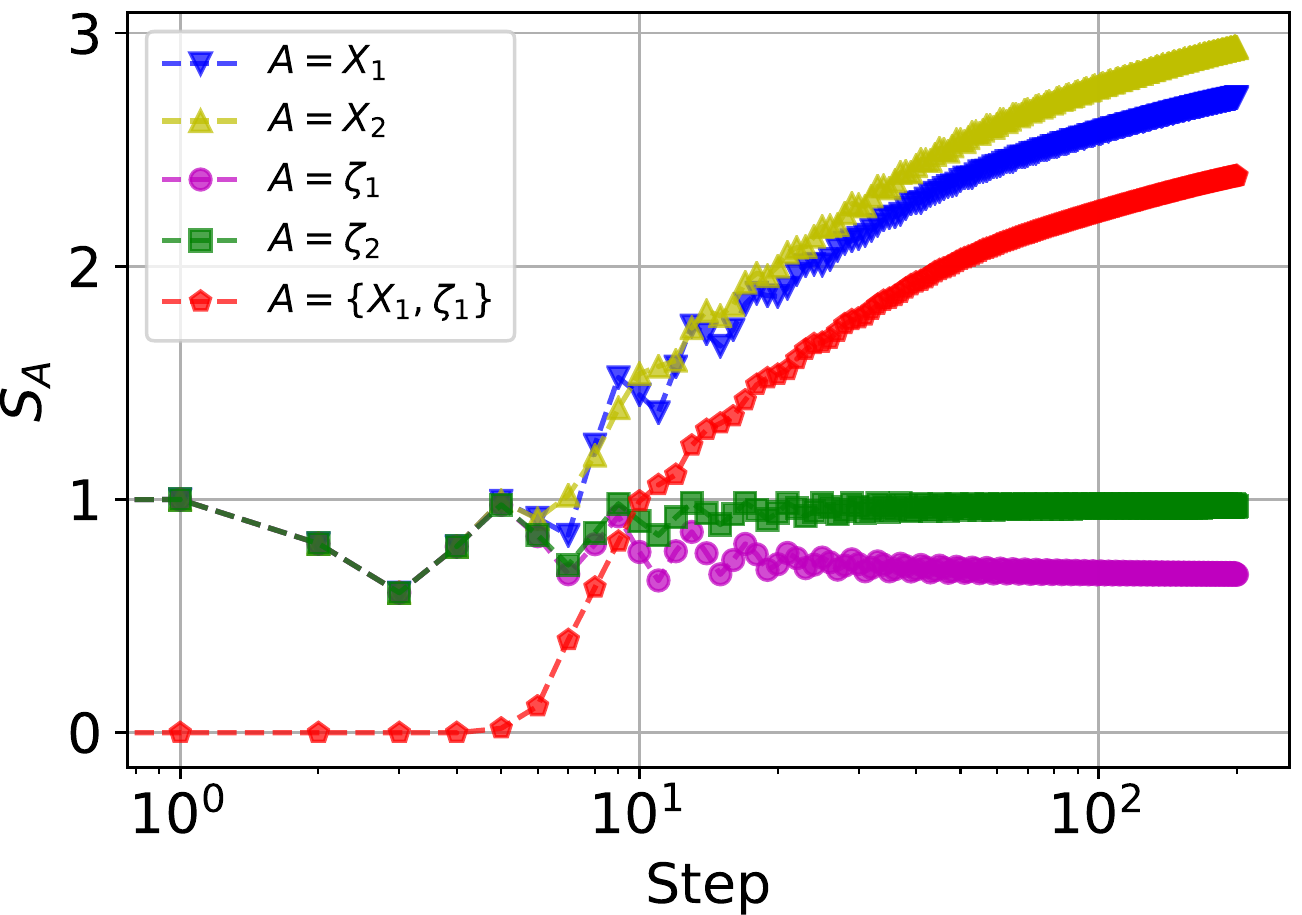}};

      \end{tikzpicture}
      
      \caption{ \label{fig:2} 
      \textit{Collision of two interacting unbiased Hadamard walks ($\mu = 3\pi/2$).} The entanglement entropy $S_A$ for various partitionings $A$. Particles 1 and 2 remain unentangled until the collision at $t = 5$ (red line), which also breaks the symmetry between the particles. The initial condition, $\ket{\Psi^{(2)}(0)} = \ket{x_{-5},+} \otimes \ket{x_5,+}$, contains the particles evenly separated by $10$ sites, but with the same spin state $\ket{+}$.}   
\end{figure}

It was shown in Ref.~\cite{Ahlbrecht_2012} that the bound states generated by the interaction exist in the band gaps of the \textit{two-body} walk operator $S^{(2)}$. While disorder generally breaks a given translational symmetry, in the ordered case $\mu_n = \mu$ the total quasi-momentum $P$ is conserved because the interaction depends only on the relative position of the particles. Therefore, the Floquet eigenstates $\ket{\phi}$ [Eq.~\eqref{eqn:Floquet}] can be enumerated by the eigenvalues $p$ of $P$ which form good quantum numbers.

We focus our attention here on the effective one-body subspace $X_1 - X_2 = 0$, and the spectrum of the projected walk operator $S^{(2)}_\mathcal{P}$, which describes the effective single-particle DTQW executed by the bound state. Substituting the plane-wave decomposition $\phi_n = \sum_p \rme^{\rmi p n } \phi_p$ into system~\eqref{eqn:treqsaux}, where $ \phi_{p} = \begin{pmatrix}
\phi_{p}^{+} & \phi_{p}^{-}
\end{pmatrix}^\mathrm{T}$ is a Bloch eigenvector, we find the dispersion relation
\begin{equation}
\label{eqn:ordereddisprel}
\cos (\mu +\phi )=\pm \frac{1}{2} \sqrt{2 \rho  \cos (2 p)-2 \rho +4},
\end{equation}
which completely determines the dynamics in the ordered case. Knowing the dispersion, the plane-wave components are given by
\begin{equation}
\frac{\phi_{p}^{+}}{\phi_{p}^{-}}
=\frac{\rme^{-\rmi \frac{3p}{2}} \sqrt{\rho } \cos (p)\pm \cos (\mu +\phi )}{\sqrt{1-\rho }}\rme^{\rmi \frac{p}{2}}.
\end{equation}

We show the dispersion relation~\eqref{eqn:ordereddisprel} in Fig.~\ref{fig:0}. In general there are two bands, whose width and gap depends on the coin parameter $\rho$. In this sense $\rho$ has the role of the hopping parameter in a standard one-dimensional tight-binding model. While the focus of the rest of this work is on the unbiased coin $\rho = 1/2$, the two maximally biased coins $\rho = 0,1$ give rise to special band structures. In particular, we find flat bands at $\mu+\phi = 0,\pm \pi$ for $\rho = 0$, and two Dirac cones in a gapless system for $\rho = 1$. In general, occurrence of flat bands results from a careful fine tuning of the tight-binding Hamiltonian~\cite{10.1088/1751-8121/aaf25c} associated with the Schrödinger equation represented by the transfer matrix.

\subsect{Symmetries of the walk}
The dynamics generated by $S^{(2)}$ has a set of important invariants. Since the shift operator $G$ moves a particle by one lattice site changing the parity of the site at each step, the single-particle walk has the sublattice symmetry $T_1 S T_1 = -S$ with $T_1 = \left(\sum_{\text{i even}} \ket{x_i}\bra{x_i} -  \sum_{\text{i odd}} \ket{x_i}\bra{x_i}\right) \otimes \mathrm{Id}_{\field{C}^2}$. For two particles, since the interaction does not change the position, the sublattice symmetry maintains the chessboard colour as the parity of the relative position $X_1 - X_2$ remains the same. Since the interaction occurs in the $X_1 - X_2 = 0$ subspace, for interactions to take place the relative position of the initial condition must therefore be chosen even as otherwise the particles never reach the interaction subspace. Generally collisions occur when $X_1 - X_2$ is non-zero and even. 

Representing $S = \rme^{-\rmi H}$ and $S^{(2)} = \rme^{-\rmi H^{(2)}}$, we can understand lattice-type symmetries through the effective lattice Hamiltonians $H$ and $H^{(2)}$. Quite formally, we may define $H^{(2)} = \rmi \ln{(S^{(2)})}$ such that the branch cut of the logarithm indicates the effective Brillouin zone for $\phi$ (the eigenvalues of $H^{(2)}$),  here taken as $-\pi < \phi \leq \pi$. When the band structure of $H$ is that of a topological insulator, this representation also underpins the ability of the quantum walk to represent the bulk-boundary phenomena and topologically protected edge states typically found in lattices with topologically non-trivial band structure~\cite{PhysRevA.82.033429}. The propagator $S$ is unitarily equivalent with alternative propagators written with a different starting point for the periodic sequence of operations constituting a step in the DTQW, termed time-frames~\cite{PhysRevB.88.121406}. Careful consideration of (the single-particle) $S$ in a symmetric time-frame allows for a consistent definition of a chiral symmetry $T_2 S T_2 = S^\dagger$ with $T_2^2 = 1$ and $T_2^\dagger = T_2$, or equivalently $T_2 H T_2 = -H$~\cite{PhysRevB.88.121406}. Eigenstates of $T_2$ then correspond to the edge states at $\phi = 0, \pi$. Similarly if $K H K = -H$, where $K$ is the complex conjugation operator, then the DTQW has a particle-hole symmetry and can represent topological phases with no counterpart in standard solid-state lattices~\cite{PhysRevB.88.121406}. However, the presence of the disordered phase interaction $V^{(2)}$ here breaks both the particle-hole and chiral symmetries. We note that the disorder also breaks the discrete time-translational symmetry of the walk. Without the disorder, the system is invariant under discrete time translations of integral multiples of the period $\tau$, and the quasi-energies are conserved upto multiples of $2\pi \hbar/\tau$.

\subsect{Entanglement entropy}
To quantify bipartite entanglement in the system we compute the von Neumann entropy of reduced density matrices. The von Neumann entropy is a good measure for bipartite entanglement because the quantum walk dynamics $S^{(2)}$ are unitary, and at any time step the system is in a pure state. For any two-fold partitioning $A, B$ of the total system $A + B$, the von Neumann entropies of the subsystems are identical, $S_A = S_B$, and given in terms of subsystem $A$ without loss of generality by
\begin{equation}
\label{eqn:SA}
S_A \equiv S(\rho_A) = -\mathrm{Tr} \left[\rho_A \log_2 \left(\rho_A \right)\right], 
\end{equation}
where $\rho_A  = \mathrm{Tr}_B \left(\rho_{AB} \right)$ and $\rho_{AB}$ is the density matrix of the total system. For a pure state $\left|{\Psi}\right\rangle_{AB} \in \mathcal{H}^{(2)}$, we have  $\rho_{AB} = \left|{\Psi}\right\rangle_{AB} \,{}_{AB}\left\langle{\Psi}\right| $.

To calculate the entanglement entropy between the two parts of the system, $A$ and $B$, we first need to define the partitioning. For example, we could take $A$ as the degree of freedom representing the position of walk 1, and then $B$ is the complementary set of degrees of freedom (position of particle 2, spin state of particle 2, and spin state of particle 1). Below, we will consider a set of different partitionings of the degrees of freedom. Evaluation of $S_A$ becomes easy in the diagonal Schmidt basis~\footnote{Our code can be found online at~\url{https://github.com/laantoi/open-science}},
\begin{equation}
\left|{\Psi}\right\rangle_{AB} = \sum_i^\chi \alpha_i \left|{\Psi_i}\right\rangle_A \otimes \left|{\Psi_i}\right\rangle_B,
\end{equation}
in terms of the real and positive Schmidt coefficients $\alpha_i$:
\begin{equation}
S_A = S_B \equiv - \sum_i^\chi \alpha_i^2 \log_2\left(\alpha_i^2\right).
\end{equation}
Here $\chi \leq \mathrm{min}\left\lbrace \mathrm{dim}\left(\mathcal{H}_A^{(1)}\right), \mathrm{dim}\left(\mathcal{H}_B^{(1)}\right) \right\rbrace$ is the Schmidt number.

\sect{Results and discussion}

We show the von Neumann entropy and probability distribution for the non-interacting case of $\mu = 0$ in Fig.~\ref{fig:1}. The non-interacting Hadamard walk on a line is well understood, with $S_A \to S_\mathrm{Had} \approx 0.872$ asymptotically for any initial state $\ket{\zeta(0)}$ if the coin is unbiased ($\rho = 1/2$) with symmetric initial states converging faster~\cite{Carneiro_2005}. Without interactions the two walks are independent, and the degrees of freedom $X_2$ and $\zeta_2$ of particle 2 do not influence the observables of particle 1, and vice versa.

\begin{figure}[b]
  \centering
        \begin{tikzpicture}
        \def\x{4.5};        \def\y{3.0};
        \def\v{-1.1};   \def\vv{-1.3};;    \def\w{10.8};	     \def\u{10.8};
			 \node at (0.0*\x,0*\y) {    \includegraphics[width=0.48	\textwidth,angle=0]{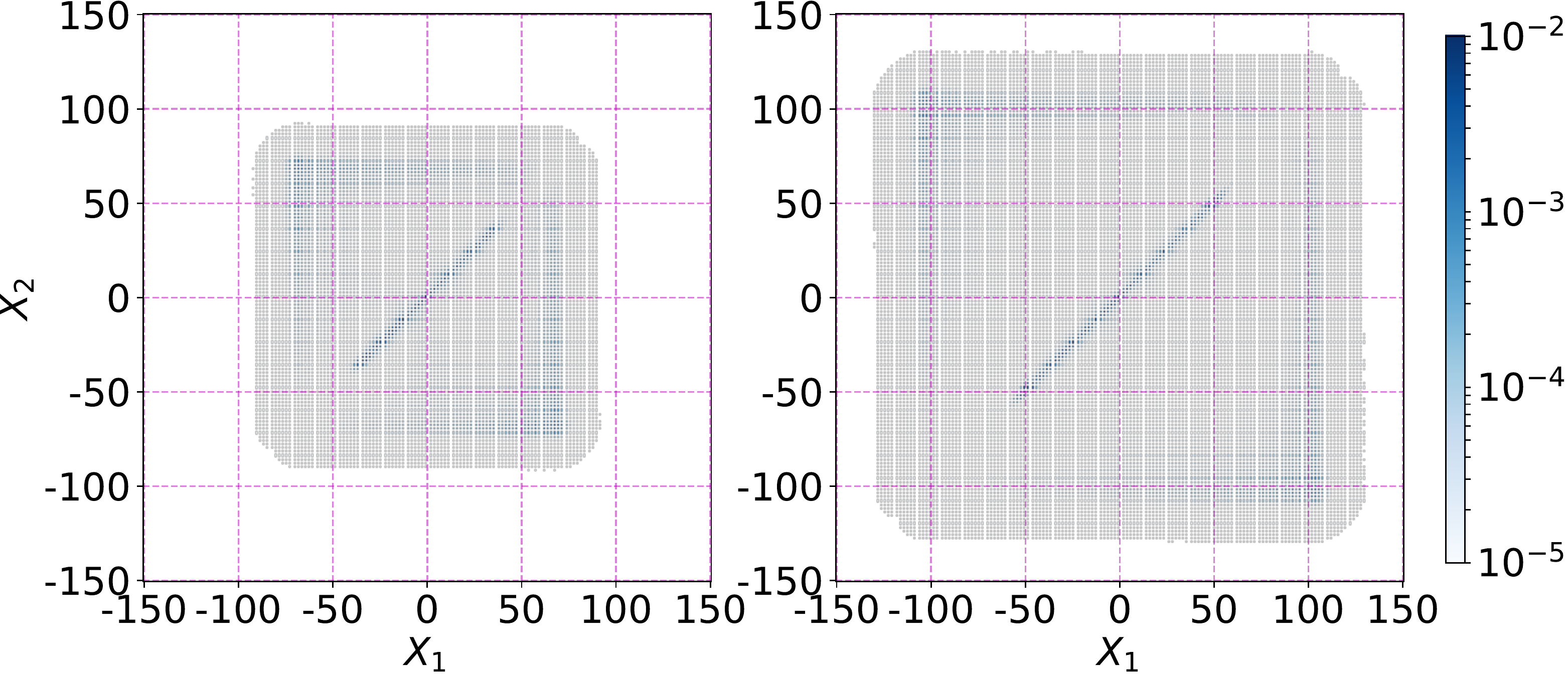}};
			\node at (-0.05*\x,-1.6*\y) {\includegraphics[width=0.4\textwidth,angle=0]{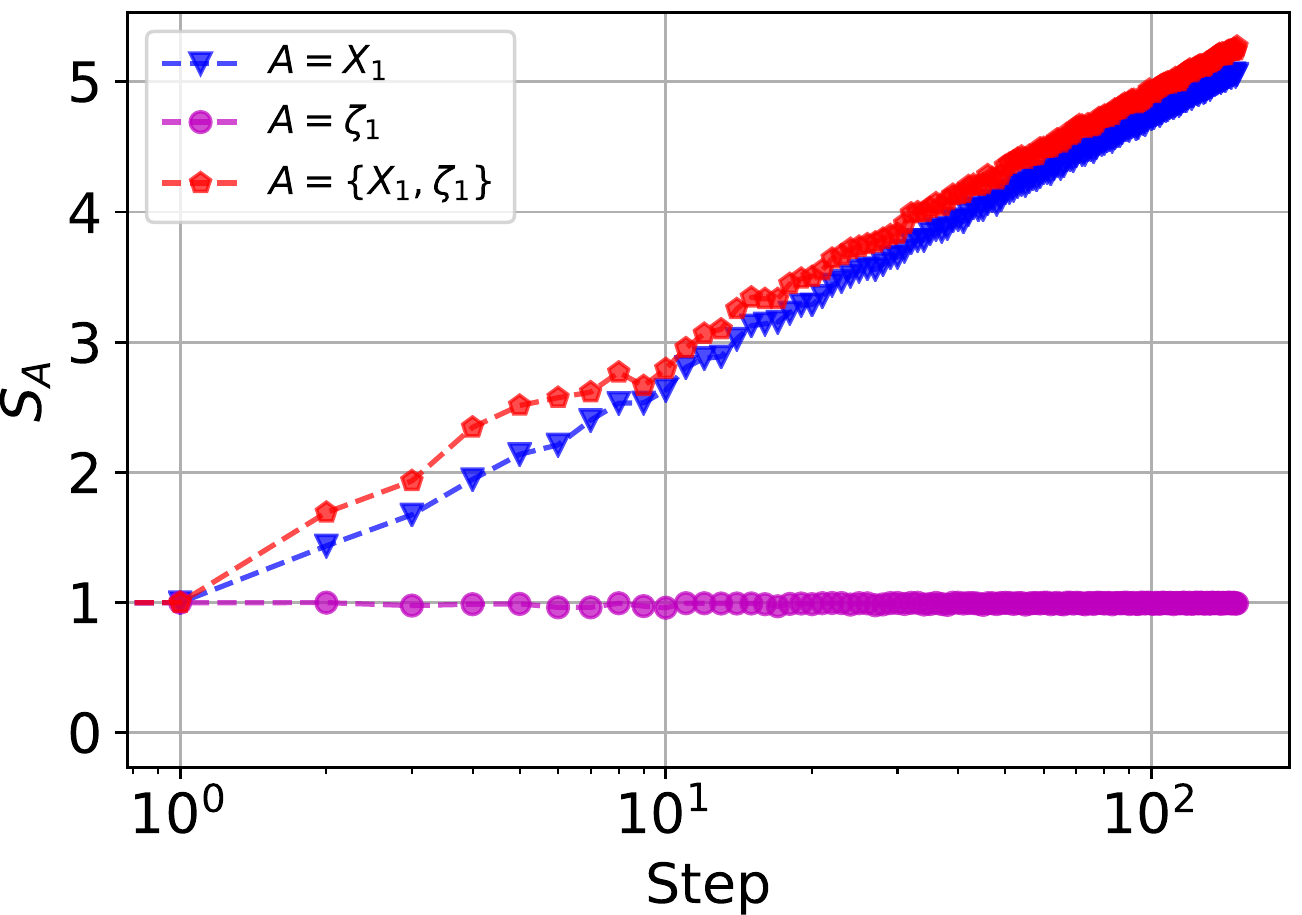}};

      \end{tikzpicture}
      
      \caption{ \label{fig:3} 
      \textit{Bound state in interacting unbiased Hadamard walks ($\mu = 3\pi/2$).} Shown is the probability distribution at $t = 100$ (top left) and $t = 150$ (top right), and the entanglement entropy (bottom). The initial condition is $\ket{\Psi^{(2)}(0)} = \ket{x_{0},-} \otimes \ket{x_0,-}$. The symmetry of the initial condition $\ket{\Psi^{(2)}(0)} = \ket{x_{0},-} \otimes \ket{x_0,-}$ imposes a similar symmetry on the partitioning meaning that the entropy $S_A$ for $A = X_2$ and $A = \zeta_2$ is the same as shown for particle 1. The entropy partitioned between the two particles (red line with pentagons) grows as $\sim \log{(t)}$ at large steps $t$.
      }   
\end{figure}

Turning on interactions without disorder, the particles become entangled upon collision, and entanglement continues to be generated in the collisional manifold $X_1-X_2 = 0$ (anti-diagonal). Starting with an initial separation of 10 sites with the same spin state, we show the entanglement dynamics in Fig.~\ref{fig:2}. Despite interaction, no bound state is formed and instead there occurs a collision at $t = 5$, which entangles the two particles. When the particles start on the anti-diagonal (same lattice site), we observe the bound state that was predicted in Ref.~\cite{Ahlbrecht_2012} (Fig.~\ref{fig:3}). That the dimer is bound is implied by the width around the anti-diagonal being exponentially localised and constant in time: around the anti-diagonal we have $P\left(\left|X_1-X_2 \right| > r \right) \sim \exp(-r)$. The centre-of-mass motion undergoes ballistic spreading ($\propto t$), but which is slower than the propagation of the single-particle peaks. The properties of the bound state without disorder are discussed in detail in Ref.~\cite{Ahlbrecht_2012}.

We assume that the disorder in $\alpha_n$ and $\beta_n$ is uniformly distributed,
\begin{equation}
 P(\theta) = \begin{cases}
 1/(2W) \qquad &\mbox{if } \epsilon - W \leq \theta \leq  \epsilon+W,\\
 0 \qquad &\mbox{otherwise},
 \end{cases}
 \end{equation} 
where $\theta \in \{\alpha_n, \beta_n\}$, and $\varepsilon$ denotes the mean of the distribution. The strength of the disorder is characterised by the width $W$ that satisfies $0 \leq W \leq \pi$ with $\pi$ representing the strongest disorder.

\begin{figure}[b]
  \centering
        \begin{tikzpicture}
        \def\x{4.5};        \def\y{-3.1};
        \def\v{-1.1};   \def\vv{-1.3};;    \def\w{10.8};	     \def\u{10.8};
			 
			 \node at (-.5*\x,0.0*\y) {    \includegraphics[width=0.23	\textwidth,angle=0]{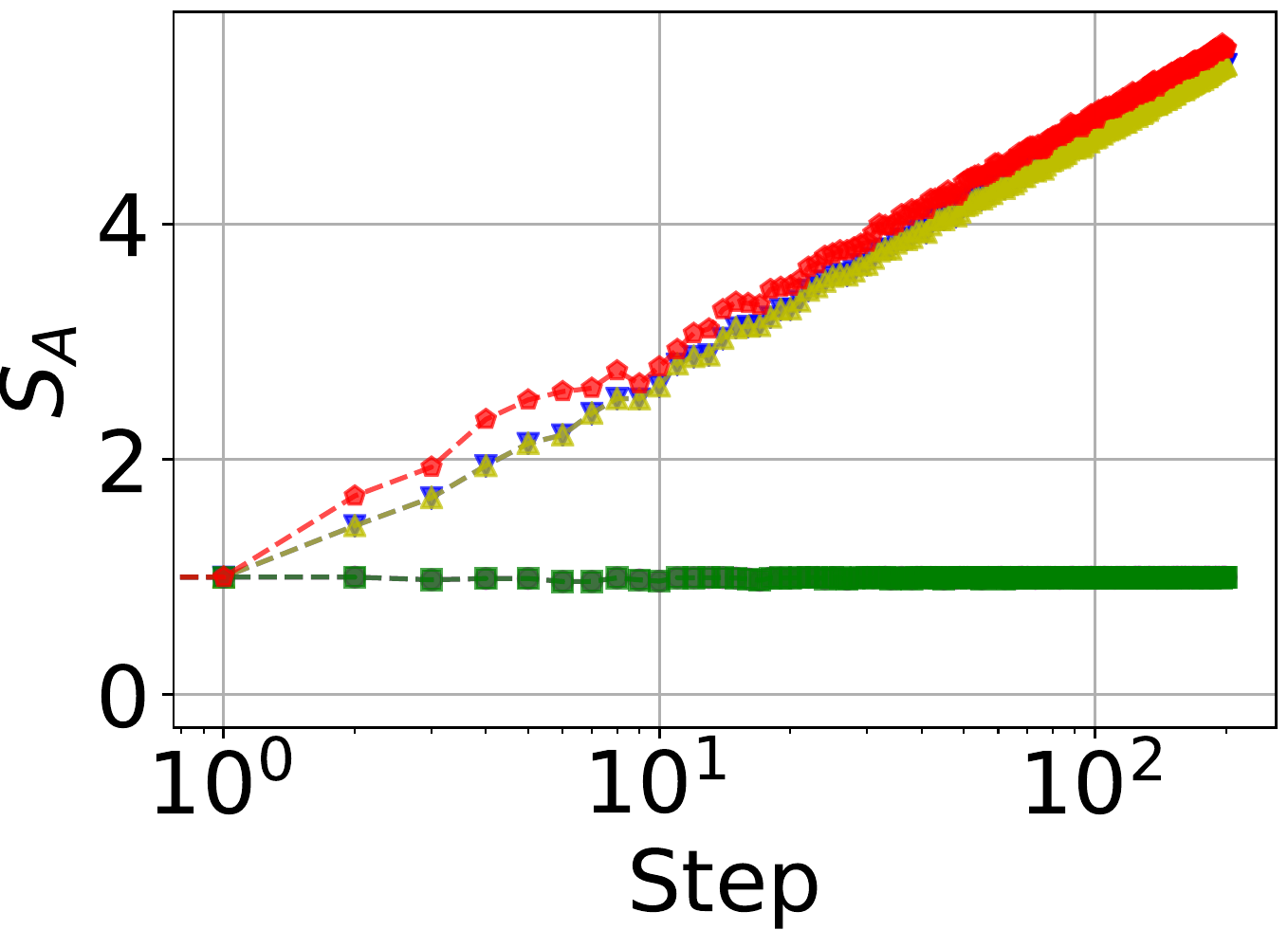}};
			 \node at (0.47*\x,0.0*\y) {    \includegraphics[width=0.23	\textwidth,angle=0]{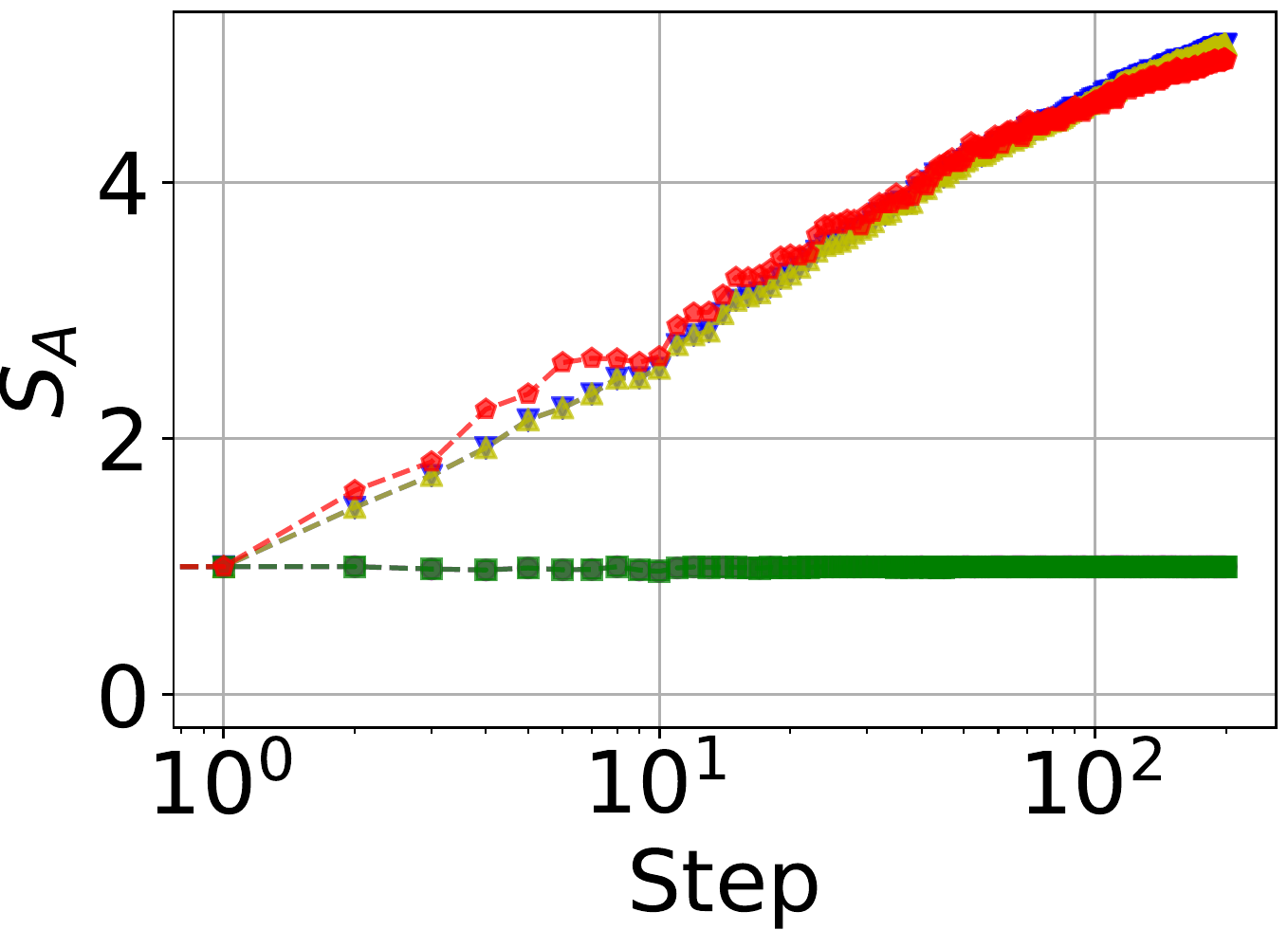}};		
			 \node at (-4.3,1.4) {(a)};			 
			 \node at (0.3,1.4) {(b)};					 

			 \node at (-.5*\x,1.0*\y) {    \includegraphics[width=0.23	\textwidth,angle=0]{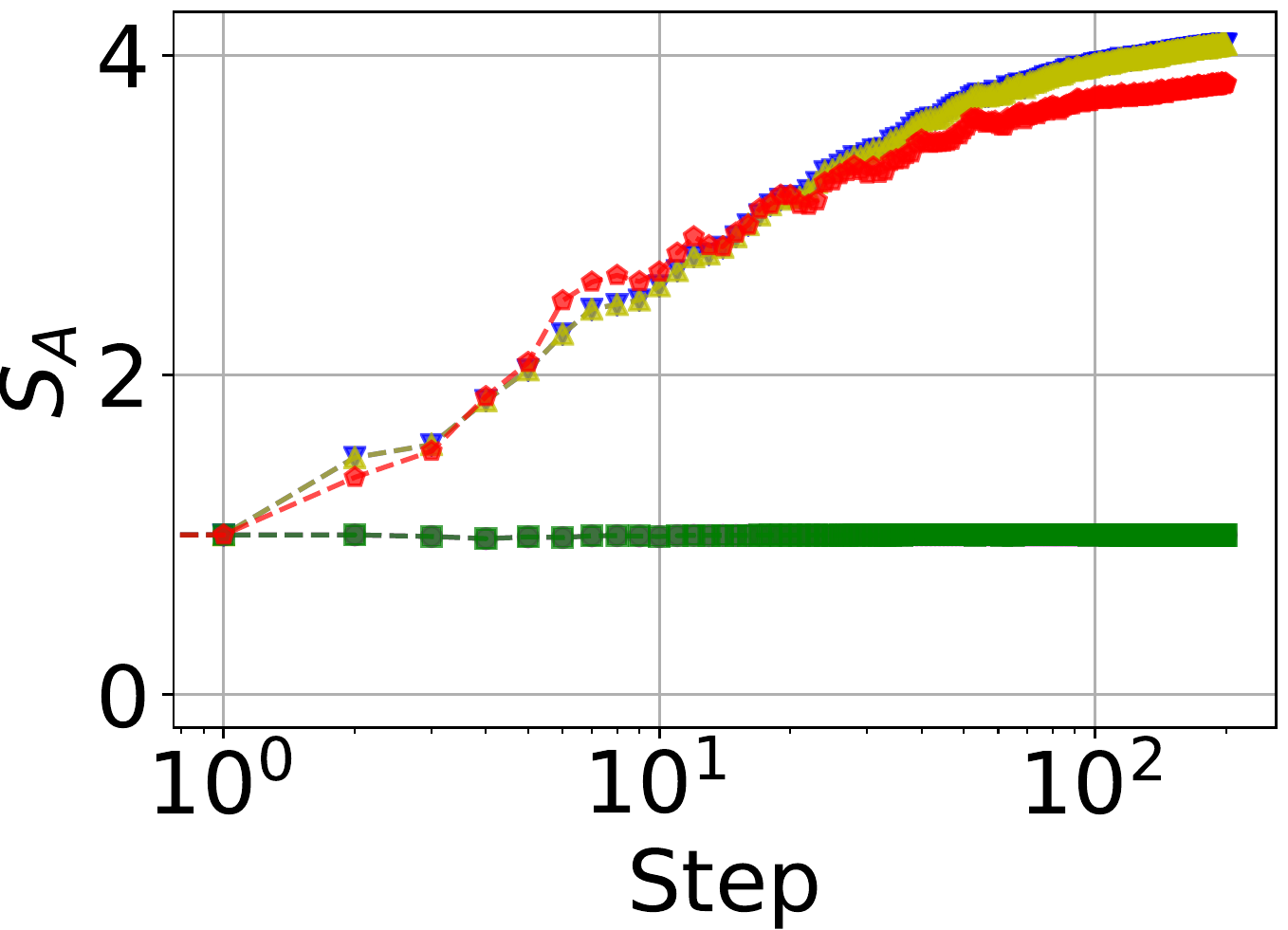}};
			 \node at (0.47*\x,1.0*\y) {    \includegraphics[width=0.23	\textwidth,angle=0]{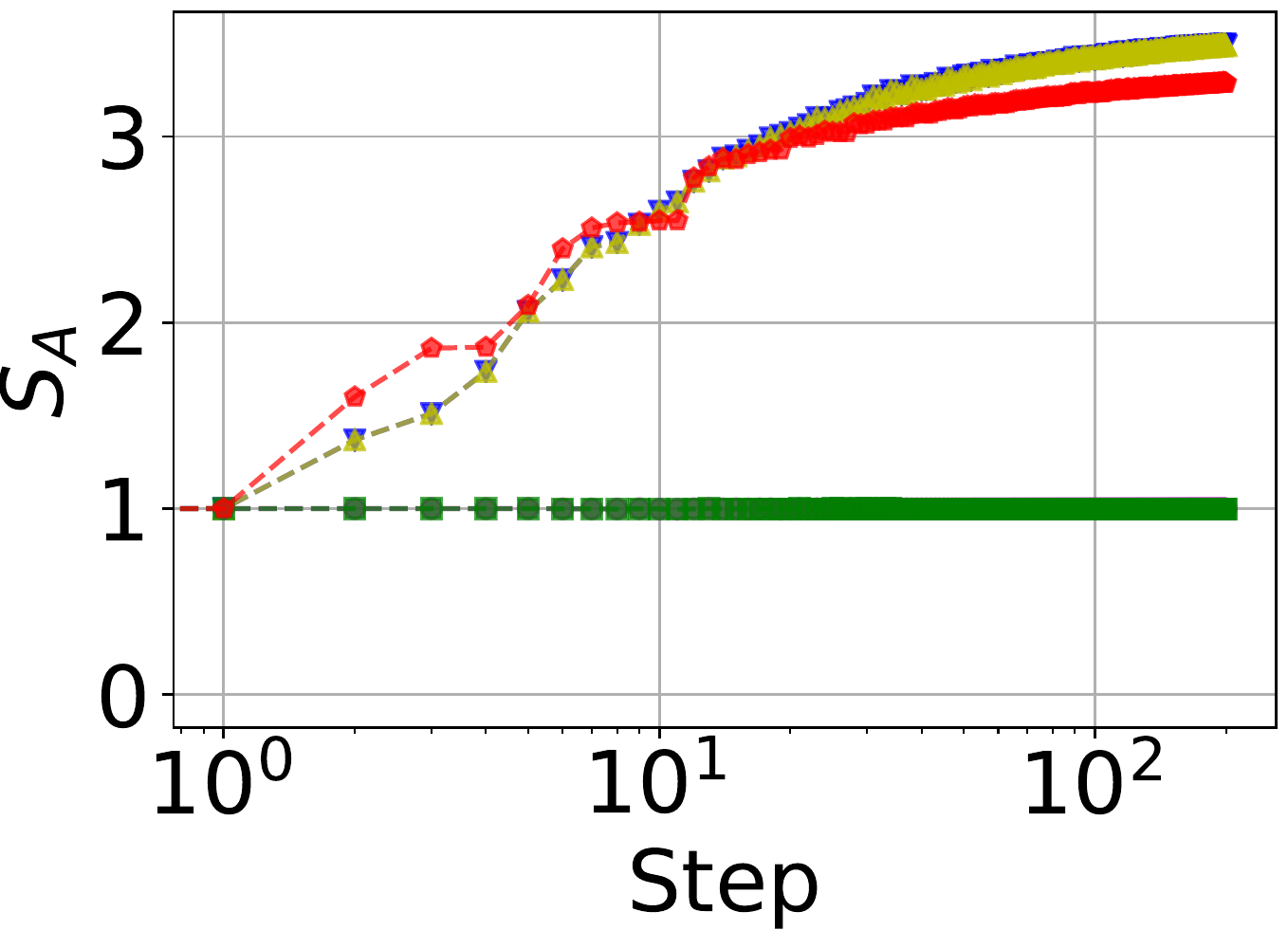}};		
			 \node at (-4.3,1.4+\y) {(c)};			 
			 \node at (0.3,1.4+\y) {(d)};

      \end{tikzpicture}
      
      \caption{ \label{fig:9} 
      \textit{Entanglement entropy vs. disorder strength $W$.} (a) $W = \pi/30$. (b) $W = \pi/6$. (c) $W = \pi/2$. (d) $W = \pi$.  The legend is the same as in Fig.~\ref{fig:2}.}
\end{figure}
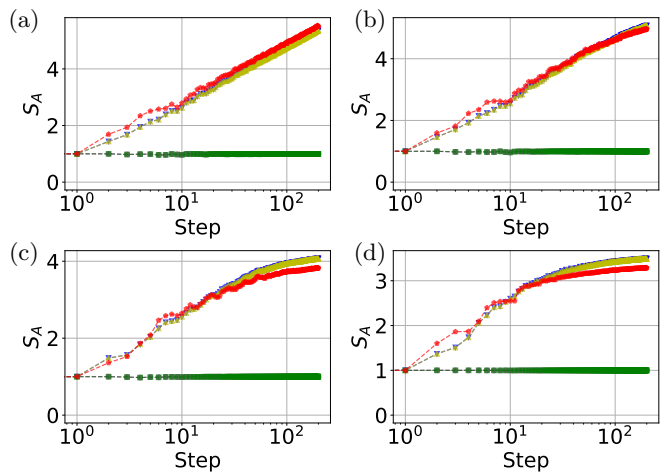  

\begin{figure}[t]
  \centering
        \begin{tikzpicture}
        \def\x{4.5};        \def\y{3.0};
        \def\v{-1.1};   \def\vv{-1.3};;    \def\w{10.8};	     \def\u{10.8};
			 \node at (0.0*\x,0*\y) {    \includegraphics[width=0.48	\textwidth,angle=0]{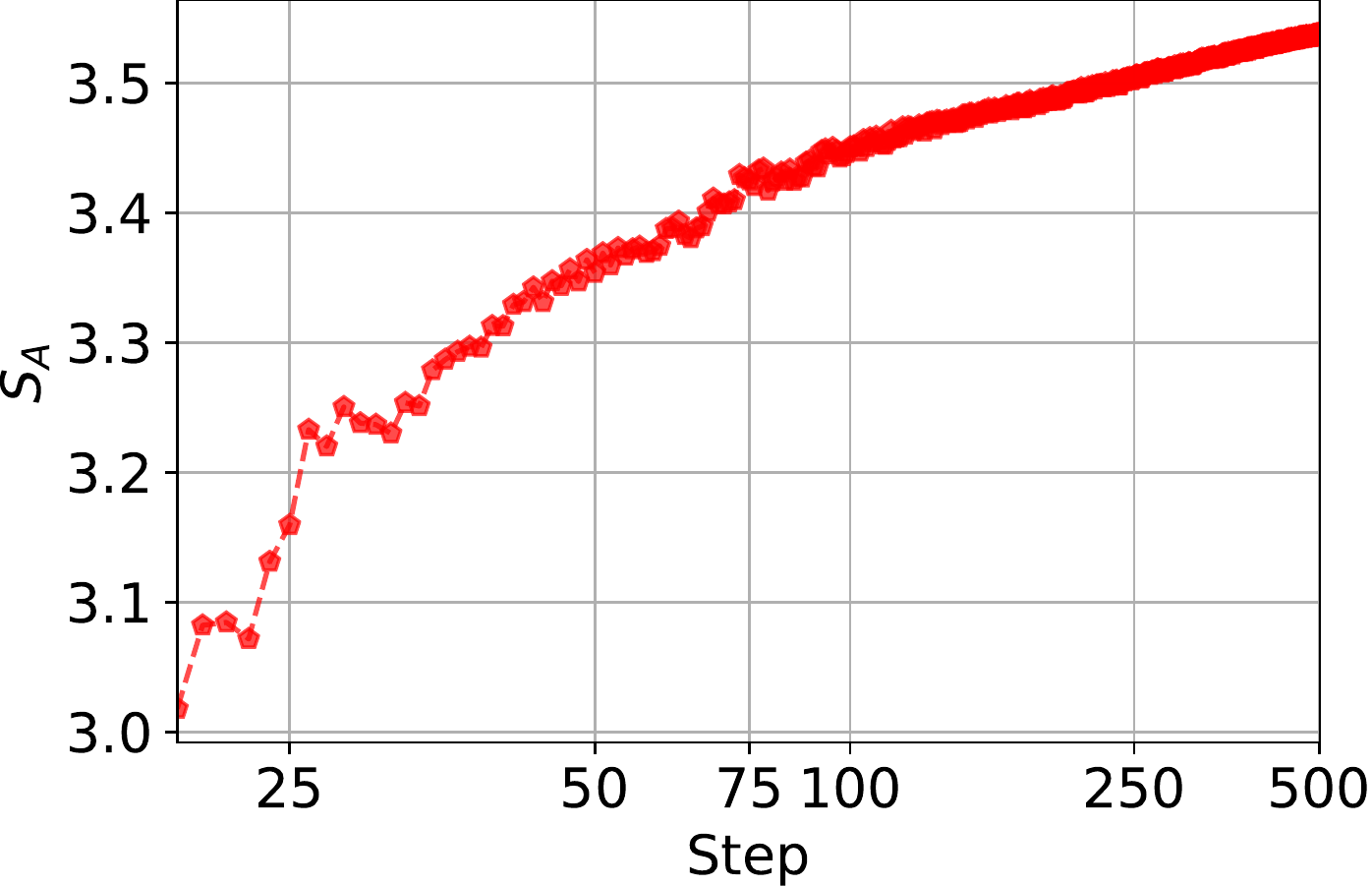}};
	
      \end{tikzpicture}
      
      \caption{ \label{fig:11} 
      \textit{Entanglement entropy with maximal disorder $W = \pi$.} 
      The entanglement grows slowly, in accordance with the double logarithmic law~\eqref{eqn:SAShannon_loc}. A straight line in this plot corresponds to a double logarithmic law as the $x$-axis is scaled by $\log(\log(t))$. The legend is the same as in Fig.~\ref{fig:2}.
      }    
\end{figure}  

\begin{figure}[b]
  \centering
        \begin{tikzpicture}
        \def\x{4.5};        \def\y{3.0};
        \def\v{-1.1};   \def\vv{-1.3};;    \def\w{10.8};	     \def\u{10.8};
			 \node at (0.0*\x,0*\y) {    \includegraphics[width=0.45	\textwidth,angle=0]{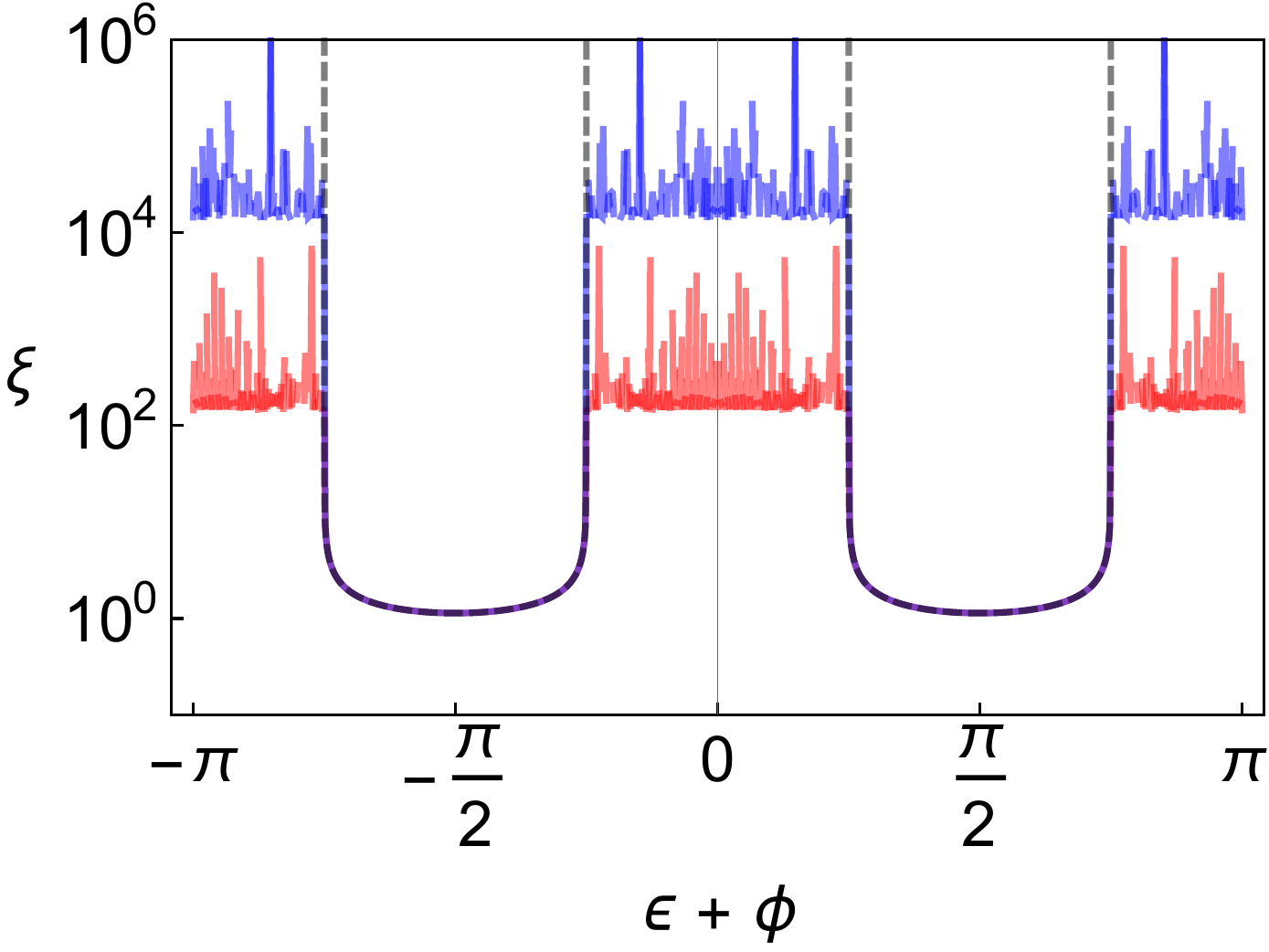}};

      \end{tikzpicture}
      
      \caption{ \label{fig:4} 
      \textit{Localisation length $\xi$ vs. quasi-energy $\phi$ without disorder.} In the numerical evaluation we use $x = 10^2$ (red) and $x = 10^4$ (blue). The dashed line is the analytical result from Eq.~\eqref{eqn:LyaAnalytic1}. We take $\alpha_n = \beta_n \equiv \varepsilon$. States in the allowed bands are delocalised, identified numerically by $\xi \gg x$ and analytically by $\Lambda_0 = 0$.}   
\end{figure}

\begin{figure}[b]
  \centering
        \begin{tikzpicture}
        \def\x{4.5};        \def\y{3.0};
        \def\v{-1.1};   \def\vv{-1.3};;    \def\w{10.8};	     \def\u{10.8};
			 \node at (0.0*\x,0*\y) {    \includegraphics[width=0.45	\textwidth,angle=0]{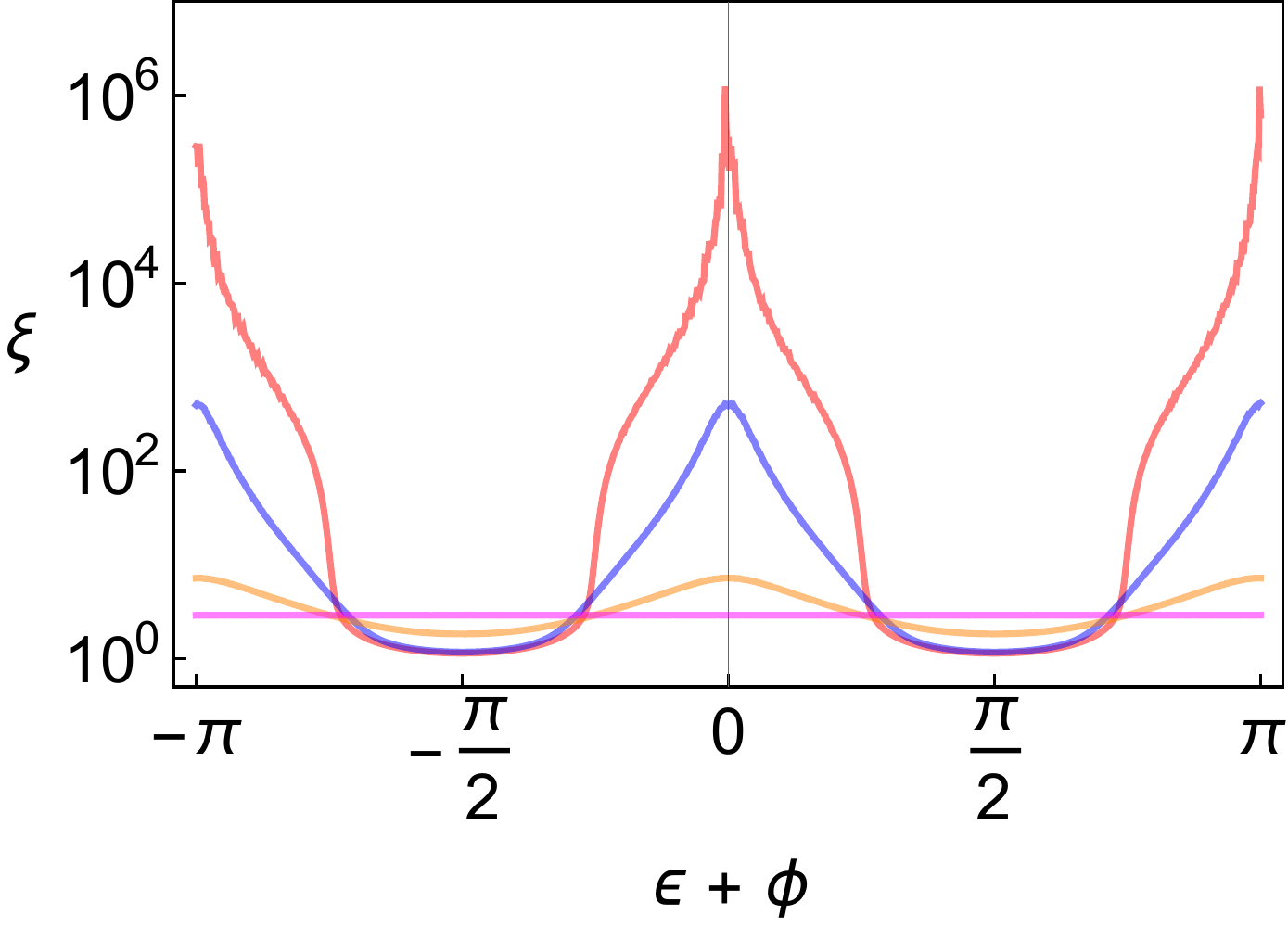}};  
      \end{tikzpicture}
      
      \caption{ \label{fig:5} 
      \textit{Localisation length $\xi$ vs. quasi-energy $\phi$ with correlated disorder.} From top to bottom at $\varepsilon + \phi = 0$: $W = \pi/30$ (red), $\pi/6$ (blue), $\pi/2$ (orange) and  $\pi$ (magenta). We take $\alpha_n = \beta_n \equiv \varepsilon$, and $x = 10^6$. The bound states of the two DTQWs become localised as a result of disorder, with a finite non-zero $\xi$. }
\end{figure}  

\begin{figure}[b]
  \centering
        \begin{tikzpicture}
        \def\x{4.5};        \def\y{3.0};
        \def\v{-1.1};   \def\vv{-1.3};;    \def\w{10.8};	     \def\u{10.8};
			 \node at (0.0*\x,0*\y) {    \includegraphics[width=0.45	\textwidth,angle=0]{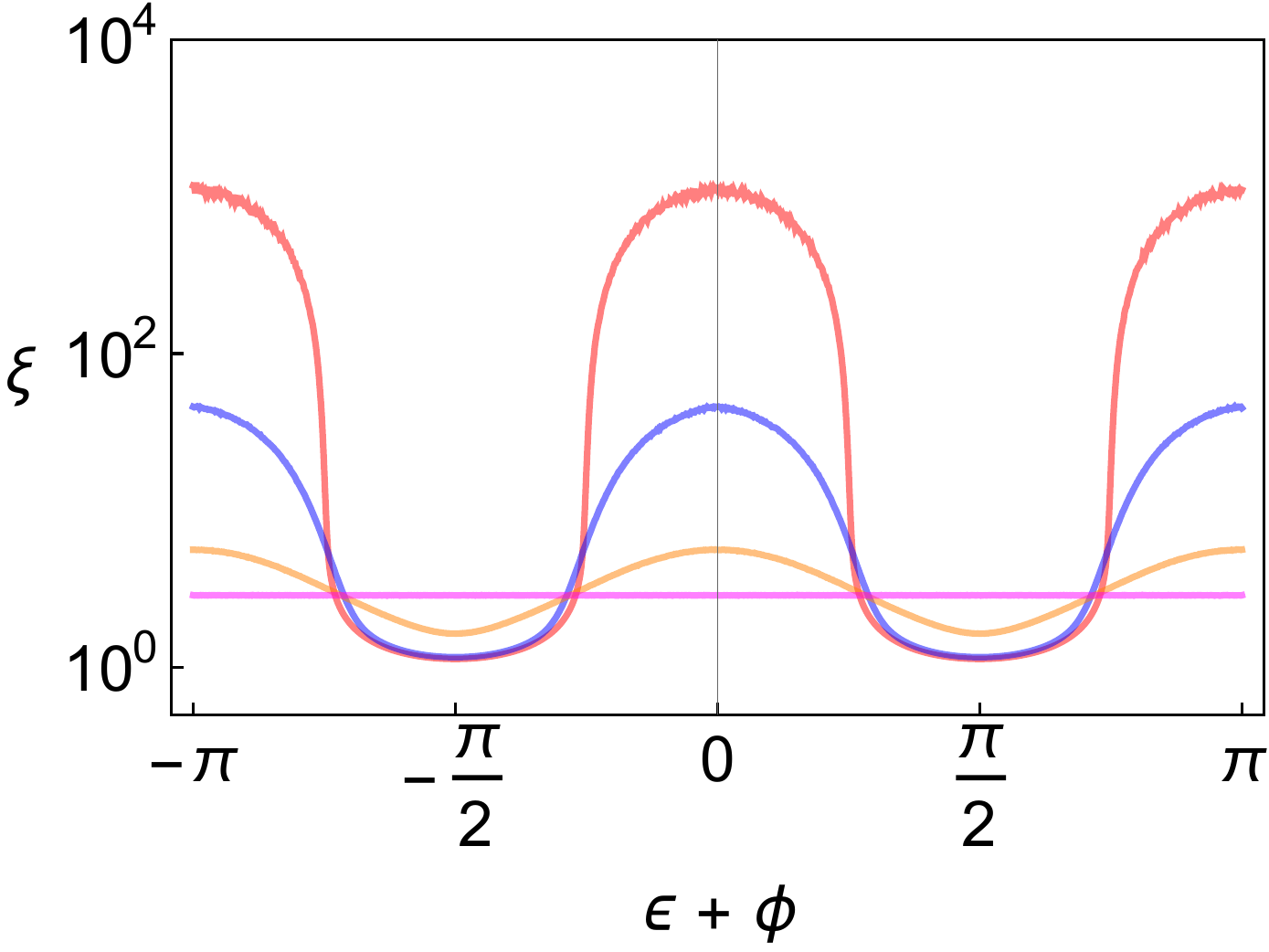}};  
      \end{tikzpicture}
      
      \caption{ \label{fig:6} 
      \textit{Localisation length $\xi$ vs. quasi-energy $\phi$ with uncorrelated disorder.} From top to bottom at $\varepsilon + \phi = 0$: $W = \pi/30$ (red), $\pi/6$ (blue), $\pi/2$ (orange) and  $\pi$ (magenta). Here $x = 10^6$, and we take $\alpha_n$ and $\beta_n$ to be uncorrelated. As a result of the uncorrelated disorder, the bound states become more strongly localised compared to the effect of correlated disorder $\alpha_n = \beta_n$ (Fig.~\ref{fig:5}), in particular, there is no distinct enhancement of $\xi$ near the band centres $\varepsilon + \phi = 0, \pm \pi$ that arises from strong correlation in the disorder in Fig.~\ref{fig:5}.}
\end{figure}

\subsect{\label{sec:3a}Disorder: Entanglent entropy}
Interparticle entanglement originates from the lack of which-path information during the interaction. In the completely delocalised ordered case ($W = 0$, $\xi \to \infty$, Fig.~\ref{fig:3} bottom), the entanglement growth for $A = \{X_1, X_2, X_1 \zeta_1 \}$ follows a logarithmic law, $S_A \sim \log(t)$. We show the effect of the disorder strength $W$ on the entanglement entropy in Fig.~\ref{fig:9}. As the strength is increased, the two-body state becomes more tightly bound perpendicular to the anti-diagonal $X_1 - X_2 = 0$, and the rate of growth of entanglement becomes slower deviating from the logarithmic law of $W = 0$. 

This can be understood from considering the projection $\mathcal{P}$ onto an effective single-particle DTQW executed by the bound state. As was shown in Ref.~\cite{Ahlbrecht_2012}, the bound state undergoes a DTQW in its own right with slower ballistic spreading: $X_1 = X_2 = \pm t/3$ for large number of steps $t$. In the single-particle case, we may write Eq.~\eqref{eqn:SA} in terms of the information theoretic Shannon entropy,
\begin{equation}
\label{eqn:SAShannon}
S_A = -\sum_i \eta_i  \log_2{\left(\eta_i \right)} = -\int_{-\infty}^\infty \d x\, \eta(x) \log_2{\left(\eta(x)  \right)},
\end{equation}
where the second equality holds only in the limit $t \to \infty$ of large number of steps. Here $\eta_i$ are the components of the density operator in the diagonal basis, $\rho_A = \sum_i \eta_i \ket{\eta_i}\bra{\eta_i}$, with $\sum_i \eta_i = 1$ giving the normalisation $\int_{-\infty}^\infty \d x\, \eta(x) = 1$. The Shannon entropy~\eqref{eqn:SAShannon} is maximised for a uniform distribution. The uniform distribution maximises the uncertainty, and the Shannon entropy can therefore be thought of as a measure of non-uniformity in the distribution. Taking the probabilities $\eta_i = \left|\Psi_i^{(2)} \right|^2$, we can therefore interpret the Shannon entropy as a measure of localisation of the quantum state with the completely delocalised phase maximising the Shannon entropy. This can be observed in Fig.~\ref{fig:3} bottom and Fig.~\ref{fig:9}: the more localised the DTQW becomes the smaller the entropy.

\begin{figure}[t]
  \centering
        \begin{tikzpicture}
        \def\x{4.5};        \def\y{3.0};
        \def\v{-1.1};   \def\vv{-1.3};;    \def\w{10.8};	     \def\u{10.8};
			 \node at (0.0*\x,0*\y) {    \includegraphics[width=0.48	\textwidth,angle=0]{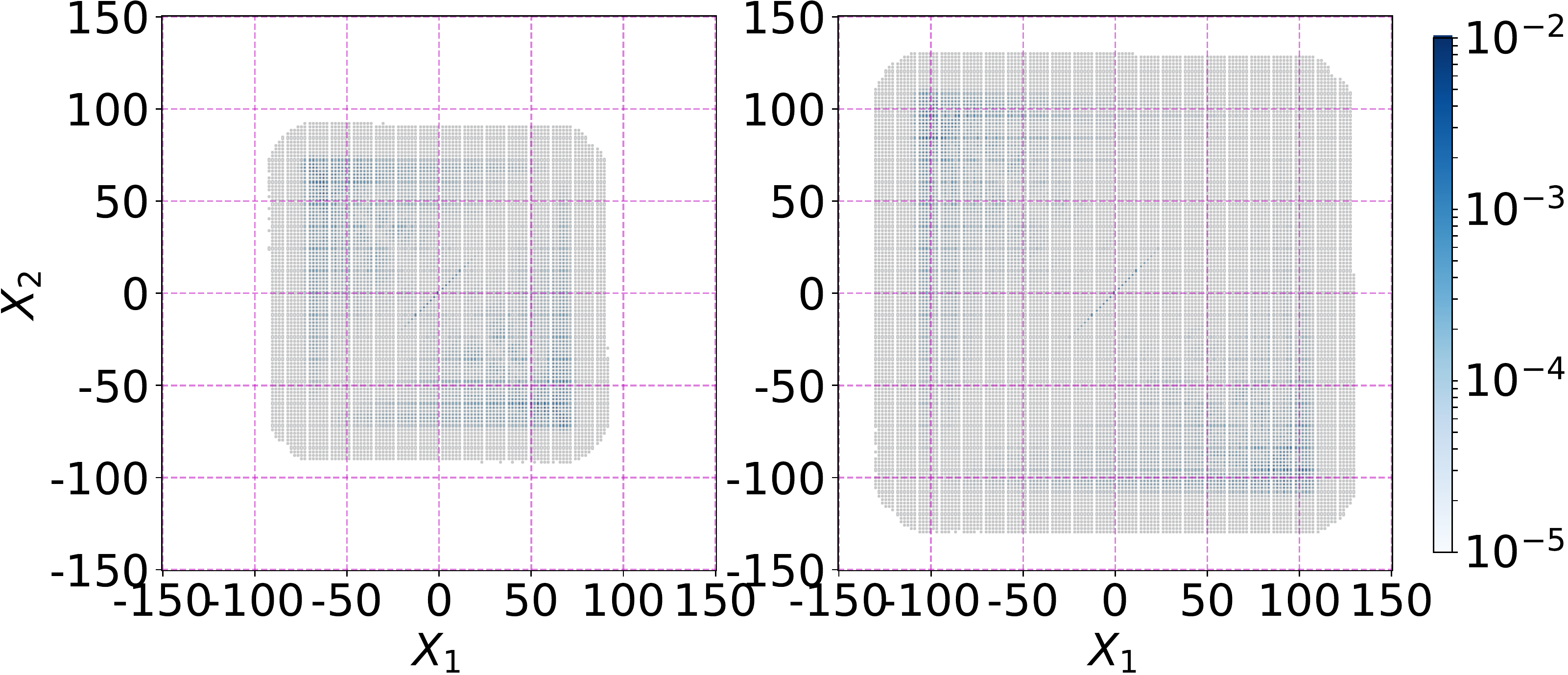}};
			\node at (0.0*\x,-1.4*\y) {    \includegraphics[width=0.43\textwidth,angle=0]{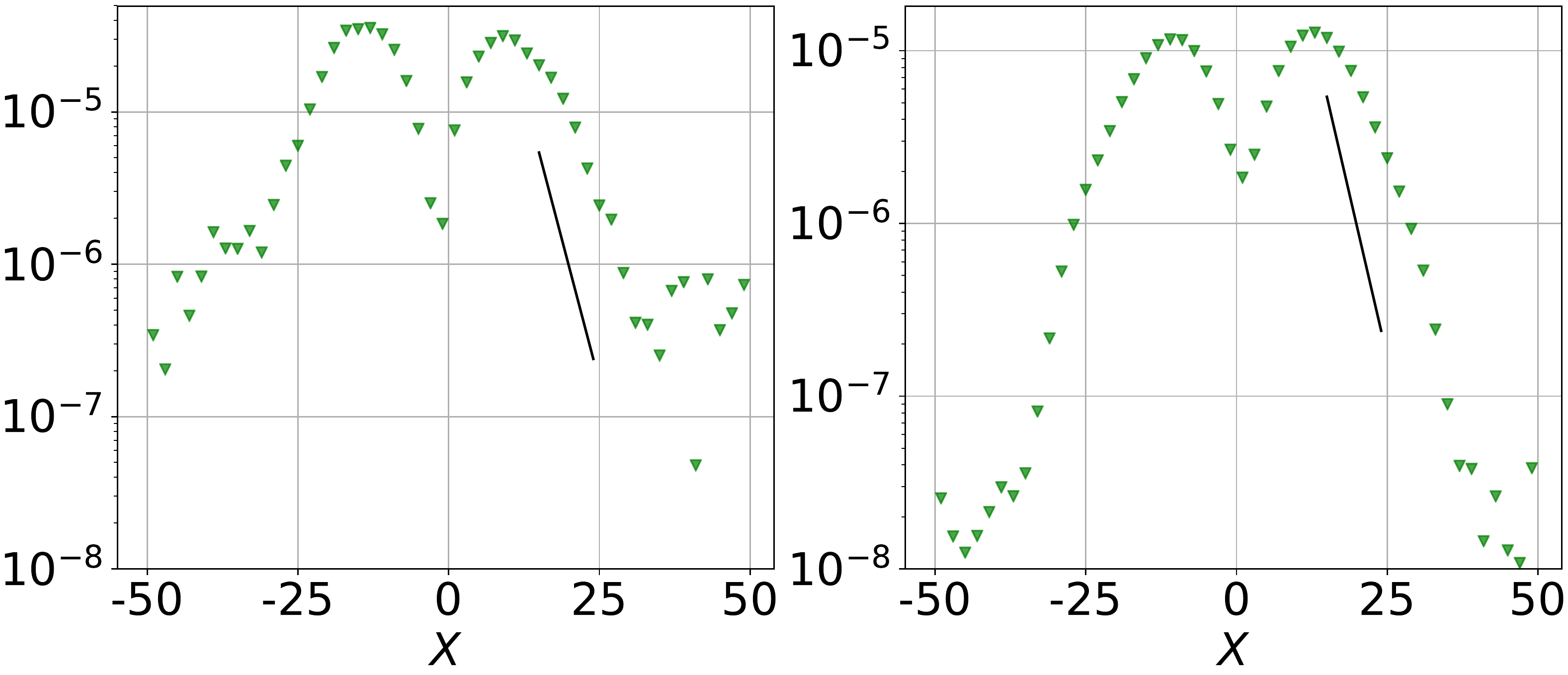}};

			 \node at (-4,2.3) {(a)};			 
			\node at (-4,-2.5) {(b)};	
      \end{tikzpicture}
      
      \caption{ \label{fig:8} 
      \textit{Localised bound state with maximal disorder $W = \pi$.} (a) The probability distribution at $t = 100$ (left) and $t = 150$ (right). (b) The density on the anti-diagonal $X_1 - X_2 = 0$ at $t = 100$ (left) and $t = 150$ (right). The solid black line is $\sim \rme^{-X/\xi}$, where $\xi = 2.886$ is the prediction from the transfer matrix approach, which agrees nicely with the localisation length obtained from direct two-body propagation (green triangles). Due to the chess board symmetry even/odd sites $X$ along the anti-diagonal are empty depending on the parity of $t$, and the empty sites are left unplotted. The initial condition is $\ket{\Psi^{(2)}(0)} = \ket{x_{0},+} \otimes \ket{x_0,+}$. 
      }   
\end{figure}

In the delocalised phase we have ballistic spreading $x \propto t$, which written in terms of the self-similar probability scaling
\begin{equation}
\eta(x,t) = \frac{1}{t^\Upsilon} \Theta\left(\frac{x}{t^\Upsilon} \right)
\end{equation}
with some exponent $\Upsilon$, gives from Eq.~\eqref{eqn:SAShannon} the logarithmic law,
\begin{equation}
\label{eqn:SAShannon_deloc}
S_A \sim \Upsilon \log{\left(t \right)}.
\end{equation}
In the localised phase, from Eq.~\eqref{eqn:AndLoc} we have $\eta_i \sim \rme^{-|i|/\xi}$, and the Shannon probabilities tend to a stationary distribution. Physically, subsequent steps of the DTQW produce exponentially little new information because the quantum state is exponentially localised, and we find a double logarithmic law
\begin{equation}
\label{eqn:SAShannon_loc}
S_A \sim \log{\left[\xi \log{(t)} \right]},
\end{equation}
where $\xi$ is the localisation length. In contrast, in the delocalised phase the ballistic spreading produces new information at each step. Similar regimes were recently found at the interface between a topological and non-topological quantum walk~\cite{10.21468/SciPostPhys.5.2.019}. We show longer-time entanglement dynamics in Fig.~\ref{fig:11}. While it is numerically challenging to propagate upto large number of steps, the observed growth of entanglement in the maximally localised phase ($W = \pi$) is in agreement with the double logarithmic law. With the maximal disorder, the localisation length is $\xi = 2.886$ (Fig.~\ref{fig:8}), and the exponentially localised envelope develops after only a relatively small number of steps. Correspondingly, we are in the large-$t$ limit for $t \gtrsim 100$, where the double logarithmic law also emerges in Fig.~\ref{fig:11}.

\subsect{\label{sec:3b}Disorder: Localisation length}

Without disorder $T_n$ becomes a constant matrix, $\mathrm{det}{\left(T_n \right)} = \lambda_+ \lambda_- = -1$ ($\alpha_n = \beta_n$), and the (maximum) Lyapunov exponent becomes
\begin{equation}
\label{eqn:LyaAnalytic1}
\Lambda_0 =   \ln{\left[  \mathrm{max} \left(\left|\lambda_+  \right|, \left|\lambda_-  \right|\right)  \right]} \geq 0
\end{equation}
in terms of the eigenvalues $\lambda_\pm$ of $T_n$. We show the localisation length in Fig.~\ref{fig:4} without disorder. Without disorder, restricting to $\alpha_n = \beta_n \equiv \varepsilon$, states inside the allowed bands of the ordered system are fully delocalised characterised by $\xi \to \infty$; the bound states consisting of the two walkers are not localised in the absence of disorder.

\begin{figure}[t]
  \centering
        \begin{tikzpicture}
        \def\x{4.5};        \def\y{3.0};
        \def\v{-1.1};   \def\vv{-1.3};;    \def\w{10.8};	     \def\u{10.8};
			 \node at (0.0*\x,0*\y) {    \includegraphics[width=0.48	\textwidth,angle=0]{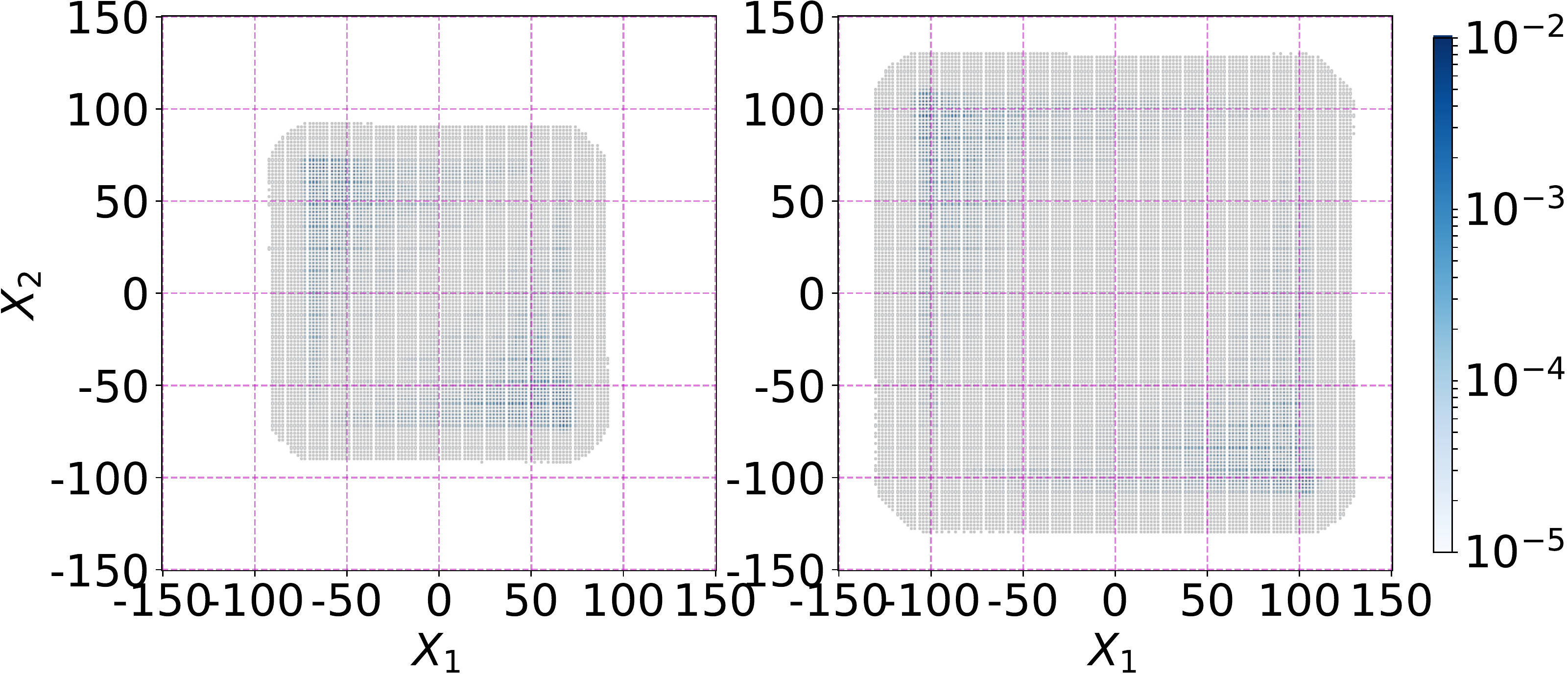}};
			\node at (-0.05*\x,-1.6*\y) {\includegraphics[width=0.4\textwidth,angle=0]{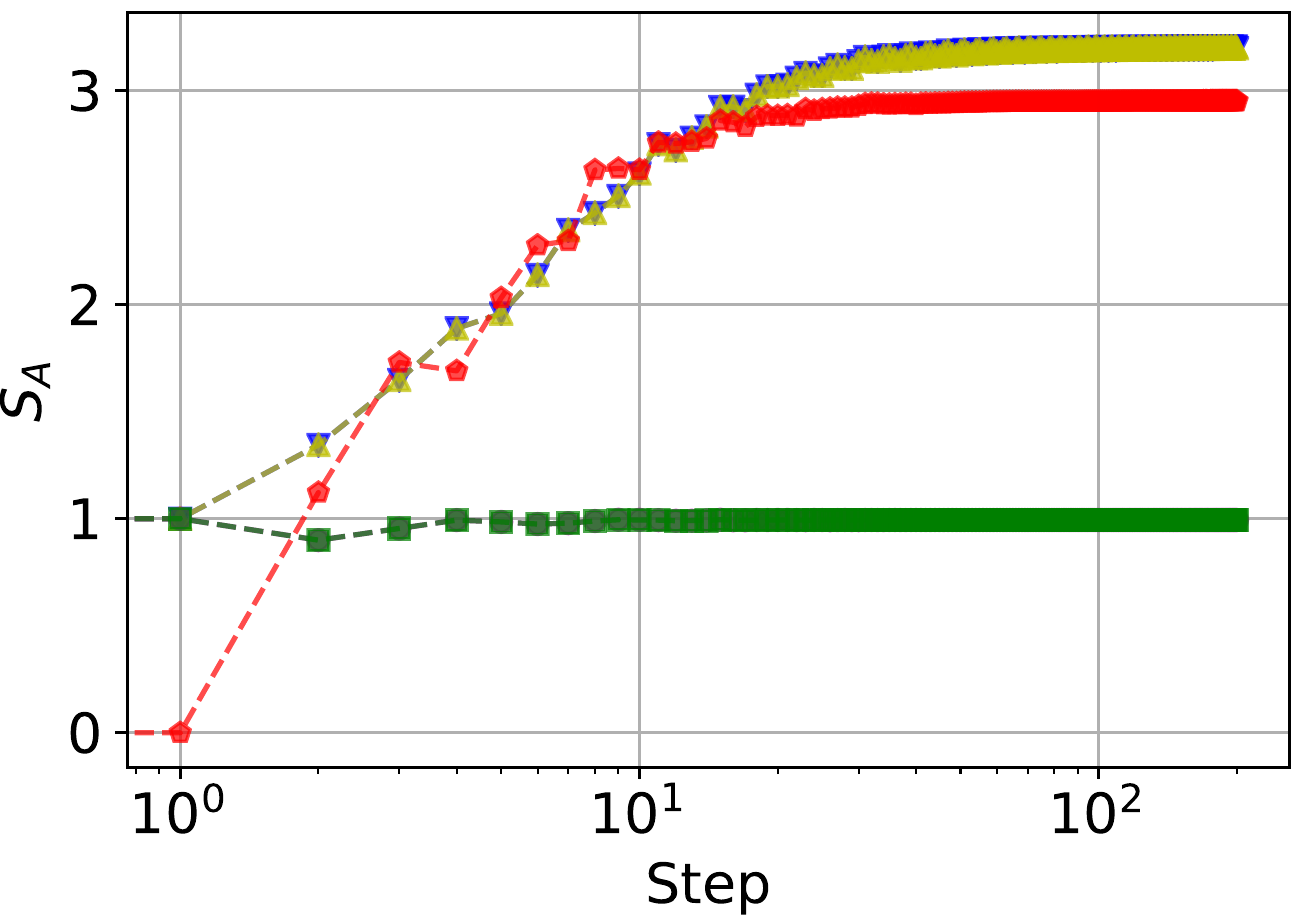}};

      \end{tikzpicture}
      
      \caption{ \label{fig:10} 
      \textit{Spin-dependent interaction in an unbiased two-body Hadamard walk ($\mu = 3\pi/2$).} Shown is the probability distribution at $t = 100$ (top left) and $t = 150$ (top right), and the entanglement entropy (bottom). The legend is the same as in Fig.~\ref{fig:2}. The bound state disappears. The initial condition is $\ket{\Psi^{(2)}(0)} = \ket{x_{0},+} \otimes \ket{x_0,+}$. 
      }   
\end{figure}

We show the effect of correlated disorder, $\alpha_n = \beta_n $, in Fig.~\ref{fig:5}. The completely delocalised states (Fig.~\ref{fig:4}) become localised with the localisation length $\xi$ being smaller the stronger the disorder $W$. We can now understand why the time-dependent disorder in the interaction $\mu(t)$ results in two-body localisation of the bound state of the two quantum walks along the anti-diagonal $X_1 - X_2 = 0$. The localisation length is finite for any non-zero disorder strength $W$, which can be understood by the physical interpretation of the diagonal disorder as a random on-site potential in a conventional Anderson system. 
We find that $\xi$ remains finite, but increases rapidly when $\varepsilon+\phi$ moves towards the centres of the allowed bands of the ordered system, and the range in the order of magnitudes spanned by $\xi$ decreases as the strength of the disorder increases. In the limit of maximal disorder, $W = \pi$, we find that $\xi$ becomes independent of $\varepsilon + \phi$. The opposite limit $W = 0$ is shown in Fig.~\ref{fig:4}. 

The same conclusions follow qualitatively from uncorrelated disorder (Fig.~\ref{fig:6}), for which the localisation is generally stronger. With correlated disorder (Fig.~\ref{fig:5}) there is a distinct enhancement of $\xi$ in the vicinity of the allowed band centres especially prominent for weak disorder, with a qualitatively different scaling with respect to $W$ in that region, which is not present with uncorrelated disorder (Fig.~\ref{fig:6}). This enhancement can be understood in terms of the strength of the disorder, which can be increased by increasing $W$ but also, where possible, by making the disorder more uncorrelated. At the band centres the weakness of the disorder is highlighted further compared to other energy scales giving rise to an anomalously large $\xi$, and together with correlations in the disorder these two effects explain the distinct enhancement of $\xi$.

We show the effect of maximal uncorrelated disorder $W = \pi$ in Fig.~\ref{fig:8}. It is clear from the probability distribution that the two-walker bound state is localised, which should be compared with the delocalised dynamics without disorder (Fig.~\ref{fig:3}). The width perpendicular to the anti-diagonal is also suppressed leading to a more tightly-bound molecule of the two walkers. The localisation length can be directly measured from the full propagation dynamics (Fig.~\ref{fig:8}b) provided it is numerically possible to propagate using a big enough system to observe the localisation; in practice we require $\xi \lesssim 5$. The measured localisation length agrees nicely (Fig.~\ref{fig:8}b) with the transfer matrix prediction $\xi = 2.886$ shown in magenta in Fig.~\ref{fig:6}. Interestingly, we find suppression of the bound state probability density around the starting point $X = 0$, which appears as part of a more general suppression along the diagonal $X_1 + X_2 = 0$.

\subsect{Spin-dependent interaction}
We note that the existence and subsequent localisation of the bound state is sensitive to the particular form of the interaction. For example, we can condition the collisional phase interaction on the spin,
\begin{equation}
\begin{split} 
V^{(2)} &= \sum_{i \in \mathbb{Z}, \zeta = \pm}  \mathrm{e}^{\mathrm{i} \zeta \mu} \left|{x_i \zeta x_i \zeta}\right\rangle \left\langle{x_i \zeta x_i \zeta}\right|  + \\
&  \sum_{\zeta_{i,j} \in \pm} \sum_{(i, j) \in \mathbb{Z}^2; i \neq j } \left|{x_i \zeta_i x_j \zeta_j}\right\rangle \left\langle{x_i \zeta_i x_j \zeta_j}\right|  ,
\end{split}
\end{equation}
in other words the interaction occurs on the subspace $X_1 - X_2 = 0$ if and only if the spin states are the same, adding a collisional phase $\pm \mu$ if the spin state is $\ket{\pm}$ respectively. The effect of the spin-dependent interaction (Fig.~\ref{fig:10}) is to act as a repulsive force between the walkers, which separate and \textit{avoid} the anti-diagonal $X_1 - X_2 = 0$, and so the generation of entanglement stops after around $100$ steps.

\sect{\label{sec:dc} Conclusions}
We have shown that disorder in the interaction between two discrete-time quantum walks results in two-body localisation of a bound state consisting of both walkers. We characterised the localisation in terms of two observables: rate of growth of the entanglement entropy, and localisation length.

We found that the bound state is robust against disorder even with maximal disorder strength $W = \pi$, and that also the weakest disorder strength $W = \pi/30$ considered results in two-body localisation. Following the interpretation in terms of the effective one-dimensional Anderson problem for the two-body bound state, even arbitrarly weak disorder should be expected to result in localisation but with a possibly large localisation length.  For the unbiased Hadamard coin $\rho = 1/2$ the localisation length in the presence of maximal disorder  is $\xi \approx 2.886$, independent of the Floquet quasi-energy. We provided numerical indications for two distinct growth laws for the entanglement: a logarithmic law in the delocalised phase and a double logarithmic law in the localised phase. Despite the absence of two-particle density transport due to localisation, and the parallels with Anderson localisation in the projected effective single-particle quantum walk, our results show that the quantum entanglement does spread albeit slowly according to the double logarithmic law. This is at variance with typical Anderson localisation, for which entanglement does not grow but instead saturates in the large $t$ limit~\cite{doi:10.1146/annurev-conmatphys-031214-014726}. On the other hand, the slow growth of entanglement is not dissimilar from many-body localised (MBL) phases~\cite{BASKO20061126}, which quench energy transport but where the entanglement exhibits slow area-law growth. 

Here, interactions and disorder are both essential ingredients to observe the two-body localisation. Without either the localisation disappears. The distinction between the ballistic expansion (thermalization) and two-body localisation is only a dynamical one, evident in the properties of the quantum states rather than in some thermodynamic or statistical measure. The Hilbert space is unbounded and without disorder the Floquet system readily absorbs `energy' from the kicks without limit (i.e. the expectation $\langle X^2 \rangle$ grows quadratically). This corresponds to ballistic expansion in real space in the quantum walk picture, and the system is ergodic exploring the entire Hilbert space. If along the anti-diagonal we truncate the real space lattice making the two-body Hilbert space bounded, the system thermalizes to essentially a uniform distribution representing infinite `temperature', where all small subsets of the whole Hilbert space are equally probable. In the Floquet case the energy is not fully conserved and the notion of temperature is not justified, but nevertheless the uniform probability distribution has completely washed out information about the initial state, and in this sense we may view it as satisfying the eigenstate thermalization hypothesis (ETH). However, as we have shown here in the presence of disorder the ETH is no longer satisfied and we may characterise the system as non-ergodic. In contrast, the system consisting of the two quantum walks remains localised in Hilbert space, retaining memory and information about the initial state, and in this sense is similar to MBL and the many-body localisation observed in Floquet systems~\cite{DALESSIO201319,PONTE2015196}.

As an outlook, an important question involves the thermodynamic limit. Is it possible to observe the localisation of more than two particles in interacting discrete-time quantum walks? Increasing the number of particles increases the amount of interaction while the strength of the disorder might be expected to play a relatively diminishing role leading possibly to a disorder-controlled transition between thermalizing and non-ergodic regimes. The MBL-like phenomena we observe here mean that the system can essentially forever remember information about the initial state, and in this sense the disorder associated with the experimental realisation with atoms in an optical lattice~\cite{Karski174} seems promising in memory components for storing quantum information. We leave for later a detailed study of how the spin configuration of the initial state affects the dynamics, and to what degree the localisation can remember and distinguish between different quantum information encoded in the initial spin states.

\acknowledgements
This work was supported by the Austrian Academy of Sciences (P7050-029-011). I would like to thank the Center for Theoretical Physics of Complex Systems at the Institute for Basic Science (Daejeon, South Korea) for hospitality, where a part of this work was carried out.

\bibliographystyle{apsrev4-1}
\bibliography{qw}

\end{document}